\documentclass[%
groupedaddress,
preprint,
showpacs,
 amsmath,amssymb,
 aps,
pra,
]{revtex4-1}

\newcommand\beq{\begin{equation}}
\newcommand\eeq{\end{equation}}

\usepackage{graphicx}
\usepackage{dcolumn}
\usepackage{bm}
\usepackage{xcolor}
\usepackage{soul}

\begin{document}


\title{Anomalous Light Transport Induced by Deeply Subwavelength Quasiperiodicity in Multilayered Dielectric Metamaterials}

\author{Marino Coppolaro}
\author{Giuseppe Castaldi}
\author{Vincenzo Galdi}
\email{vgaldi@unisannio.it}
\affiliation{Fields \& Waves Lab, Department of Engineering, University of Sannio, I-82100 Benevento, Italy}

\date{\today}


\begin{abstract}
For dielectric multilayered metamaterials, the effective-parameter representation is known to be insensitive to geometrical features occurring at deeply subwavelength scales. However, recent studies on periodic and aperiodically ordered geometries have shown the existence of certain critical parameter regimes where this conventional wisdom is upended, as the optical response of finite-size samples may depart considerably from  the predictions of standard effective-medium theory. In these regimes, characterized by a mixed evanescent/propagating light transport, different classes of spatial (dis)order have been shown to induce distinctive effects in the optical response, in terms of anomalous transmission, localization, enhancement, absorption and lasing.
Here, we further expand these examples by considering a quasiperiodic scenario based on a modified-Fibonacci geometry. Among the intriguing features of this model there is the presence of a scale parameter that controls the transition from perfectly periodic to quasiperiodic scenarios of different shades. Via an extensive parametric study, this allows us to identify the quasiperiodicity-induced anomalous effects, and to elucidate certain distinctive mechanisms and footprints.  
Our results hold potentially interesting implications for the optical probing of structural features at a resolution much smaller than the wavelength, and could also be leveraged to design novel types of absorbers and low-threshold lasers.
\end{abstract}

\maketitle

\section{Introduction}

Dielectric multilayers constitute one of the simplest and most common classes of optical metamaterials \cite{Capolino:2009vr,Cai:2010om}, and can be fabricated with high precision via well-established deposition processes. In the regime of {\em deeply subwavelength} layers, spatial-dispersion (nonlocal) effects tend to be negligibly weak, so that these materials can be accurately modeled via
macroscopic effective parameters that do not depend on the specific geometrical order and thickness of the layers, but only on their constitutive properties and filling fractions.
This {\em effective-medium-theory} (EMT) model \cite{Sihvola:1999em} is known to capture the macroscopic optical response quite accurately. However, recent theoretical \cite{Sheinfux:2014sm} and experimental studies \cite{Zhukovsky:2015ed} on periodic arrangements have pointed out that nonlocal effects may be counterintuitively amplified within certain critical parameter regimes mixing evanescent and propagating light transport, thereby leading to the {\em breakdown} of the EMT approximation. Follow-up studies \cite{Andryieuski:2015ae,Popov:2016oa,Lei:2017rt,Maurel:2018so,Castaldi:2018be,Gorlach:2020bc} have provided alternative interpretations of these effects, and have suggested possible  corrections to the conventional EMT model in order to capture them. These corrections typically include frequency- and wavenumber-dependent terms to account for nonlocality, and possibly magneto-electric coupling to ensure self-consistency.
In essence, the above results indicate that the optical response of finite-size, fully dielectric multilayered metamaterials may exhibit an anomalous sensitivity to geometrical features at deeply subwavelength scales, which may find intriguing applications in numerous fields, ranging from optical sensing to switching and lasing.

A fascinating and substantially uncharted implication of the above outcomes is that spatial order (or disorder) may play a key role not only in the {\em diffractive} regime of wavelength-sized layers (typical, e.g, of photonic crystals \cite{Joannopoulos:2008pc}), but also at much smaller scales. For instance, theoretical \cite{Sheinfux:2016cr} and experimental \cite{Sheinfux:2017oo} studies in {\em randomly disordered} dielectric multilayers have demonstrated the occurrence of anomalous Anderson-type localization effects in stark contrast with the EMT prediction of an essentially transparent behavior. Within this framework, we have recently initiated a systematic exploration of {\em aperiodically ordered}  geometries \cite{Macia:2006tr,DalNegro:2011da}, which constitute the middle ground between perfect periodicity and random disorder. These geometries have been extensively studied in the diffractive regime of photonic ``quasicrystals'' 
\cite{Poddubny:2010pc,Vardeny:2013hj,Ghulinyan:2014od}, but their interplay with mixed evanescent/propagating light transport at deeply subwavelength scales remains largely unexplored. In particular, we have studied the Thue-Morse \cite{Coppolaro:2018ao} and Golay-Rudin-Shapiro  \cite{Coppolaro:2020eo} geometries, characterized by {\em singular-continuous} and {\em absolutely continuous} spatial spectra, respectively \cite{Grimm:2015ac}; from a measure-theoretic viewpoint (Lebesgue decomposition theorem), these represent two of the three distinctive spectral traits of aperiodic order \cite{Grimm:2015ac}. 
For these geometries, we have explored the critical parameter regimes leading to the occurrence of the EMT-breakdown phenomenon, highlighting some similarities and fundamental differences from what observed in the periodic and random scenarios.

To close the loop, here we focus on  {\em quasiperiodic} geometries characterized by {\em discrete} spatial spectra, representing the remaining of the aforementioned distinctive traits \cite{Grimm:2015ac}, which has never been explored in connection with deeply subwavelength dielectric multilayers.  In this context, the quintessential representative geometries are based on Fibonacci-type sequences \cite{Albuquerque:2004pa}. Specifically, here we consider a
modified-Fibonacci geometry \cite{Buczek:2005pd} characterized by a scale-ratio parameter that can be exploited to study the transition from periodic to quasiperiodic order, so as to identify and elucidate the  anomalous light-transport effects genuinely induced by quasiperiodic order.

Accordingly, the rest of the paper is organized as follows. In Sec. \ref{Sec:Formulation}, we outline the problem and describe its geometry and main observables. In Sec. \ref{Sec:Results},
we illustrate some representative results from a comprehensive parametric study, indicating the occurrence of anomalous light-transport effects (in terms of transmittance, field enhancement, absorption, and lasing) that are in striking contrast with the predictions from conventional EMT and with what observable in periodic counterparts. We also address the development of nonlocal corrections that can capture some of these effects.
Finally, in Sec. \ref{Sec:Conclusions}, we draw some conclusions and outline some possible directions for further research.

\section{Problem Formulation}
\label{Sec:Formulation}

\subsection{Geometry}
\begin{figure}
	\centering
	\includegraphics[width=10cm]{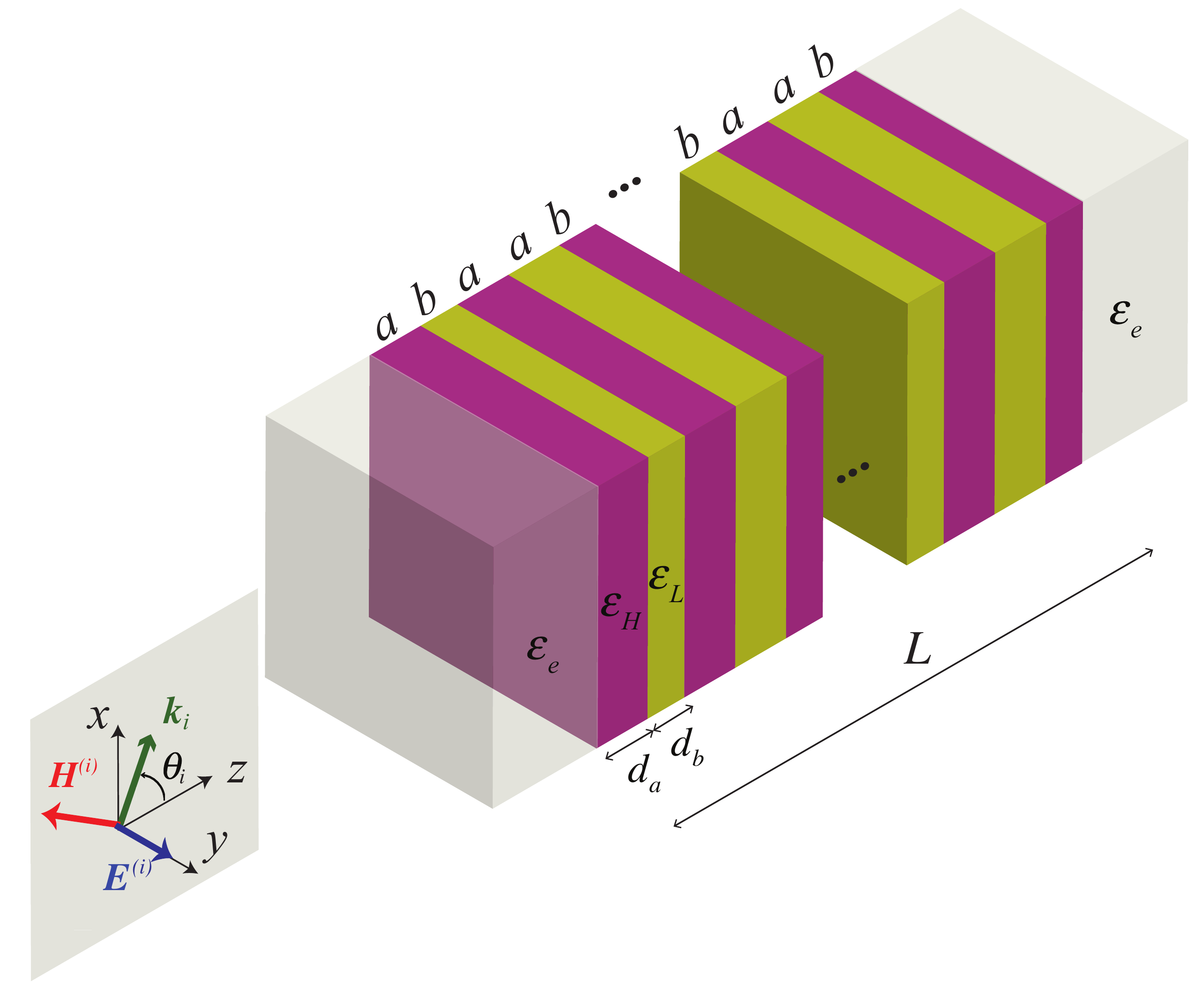}
	\caption{Problem schematic: A dielectric multilayered metamaterial with modified-Fibonacci geometry (details in the text), embedded in a homogeneous dielectric background with relative permittivity $\varepsilon_e$,  is obliquely illuminated by a plane wave with TE polarization.}
	\label{Figure1}
\end{figure}

The geometry of interest is schematically illustrated in Fig. \ref{Figure1}. Our multilayered metamaterial is composed of dielectric layers with alternating high and low relative permittivity ($\varepsilon_H$ and $\varepsilon_L$, respectively), and generally different thicknesses $d_a$ and $d_b$ distributed according to the Fibonacci sequence. The structure is assumed of infinite extent along the $x$ and $y$ directions, and is embedded in a homogeneous background with relative permittivity $\varepsilon_e$. We assume that all materials are nonmagnetic (relative permeability $\mu=1$) and, for now, we neglect optical losses.

The quasiperiodic Fibonacci geometry can be equivalently generated in several ways. One possibility is to iterate the well-known inflation rules \cite{Albuquerque:2004pa}
\beq
a \rightarrow a b, \quad b \rightarrow a,
\label{eq:itrule}
\eeq
associating the thicknesses $d_a$ and $d_b$ to the symbols $a$ and $b$, respectively, in the obtained sequence. Equivalently, one can exploit a {\em cut-and-project} approach, and calculate directly the positions of the layer interfaces as \cite{Buczek:2005pd}
\beq
z_{n}=d_a\left\|\frac{n}{\varphi}\right\|+d_b\left(n-\left\|\frac{n}{\varphi}\right\|\right),
\label{eq:Fibseq}
\eeq
where $\varphi \equiv(1+\sqrt{5}) / 2\approx1.618$ is the Golden Mean and 
\beq
\|x\|=\left\{\begin{array}{ll}
	n, & n \leq x<n+\frac{1}{2},\\
	n+1, & n+\frac{1}{2} \leq x \leq n+1.
\end{array}\right.
\eeq
It can be shown that, in the asymptotic limit of an infinite sequence, the ratio between the number of symbols  $a$ and $b$ approaches the Golden Mean \cite{Buczek:2005pd} , viz.,
\beq
\lim _{N \rightarrow \infty} \frac{N_{a}}{N_{b}}=\varphi, \quad N_{a}+N_{b}=N.
\label{eq:aslim}
\eeq
It is important to note that, at variance with typical Fibonacci-type multilayer geometries in the literature 
\cite{Vasconcelos:1999tf}, here we only assume the layer thicknesses distributed
according to the Fibonacci sequence, whereas the relative permittivities are simply alternated; this implies that, for each layer, there are four possible combinations of thickness and relative permittivity. This modified scheme facilitates the comparison with the EMT predictions as well as with a periodic reference structure. Accordingly, we generally assume $d_a\ge d_b$, and define the scale-ratio parameter
\beq
\nu=\frac{d_b}{d_a}, \quad 0<\nu \leq 1.
\label{eq:nu}
\eeq
By changing $\nu$, we can study the transition between perfect periodicity ($\nu=1$) and variable shades of quasiperiodic order ($\nu<1$). Within this framework, it is expedient to define the average layer thickness $\bar{d}=L/N$, with $L$ denoting the total thickness of the multilayer (see Fig. \ref{Figure1}). By exploiting the result in (\ref{eq:aslim}), it can be readily shown that, in the asymptotic limit of an infinite sequence,
\beq
\bar{d}=\frac{\varphi d_a+d_b}{1+\varphi}.
\label{eq:dav}
\eeq

As previously mentioned, the spatial spectrum associated with our modified-Fibonacci geometry is discrete \cite{Albuquerque:2004pa}. Specifically, it can be shown that, in the asymptotic limit of an infinite sequence, there is a double infinity of spectral peaks localized at wavenumbers \cite{Buczek:2005pd}
\beq
k_{zpq}=\frac{2 \pi}{\bar{d}} \frac{\left(p+q \varphi\right)}{(\varphi+1)},
\label{eq:kzpq}
\eeq
with amplitudes
\beq
S_{pq}=\frac{\sin W_{pq}}{W_{pq}}, 
\eeq
where
\beq
W_{pq}=\frac{\pi}{\bar{d}}\left(p d_a-q d_b\right)=\frac{\pi(1+\varphi)\left(p-q \nu\right)}{\nu+\varphi}.
\label{eq:Wpq}
\eeq
As typical of quasiperiodicity, the above spectrum is generally characterized by pairwise-incommensurate harmonics \cite{Buczek:2005pd}. Quite interestingly, it can be shown \cite{Buczek:2005pd} (see also Appendix \ref{Sec:AppA} for details) that, for commensurate scales (i.e,. rational values of the scale ratio), the spatial spectrum is periodic, even though the geometry remains aperiodic in space. Moreover, it can be verified that for the periodic case ($\nu=1$, i.e., $d_a=d_b$), the conventional periodic spatial spectrum is recovered (see  Appendix \ref{Sec:AppA} for details).

\begin{figure}
	\centering
	\includegraphics[width=16cm]{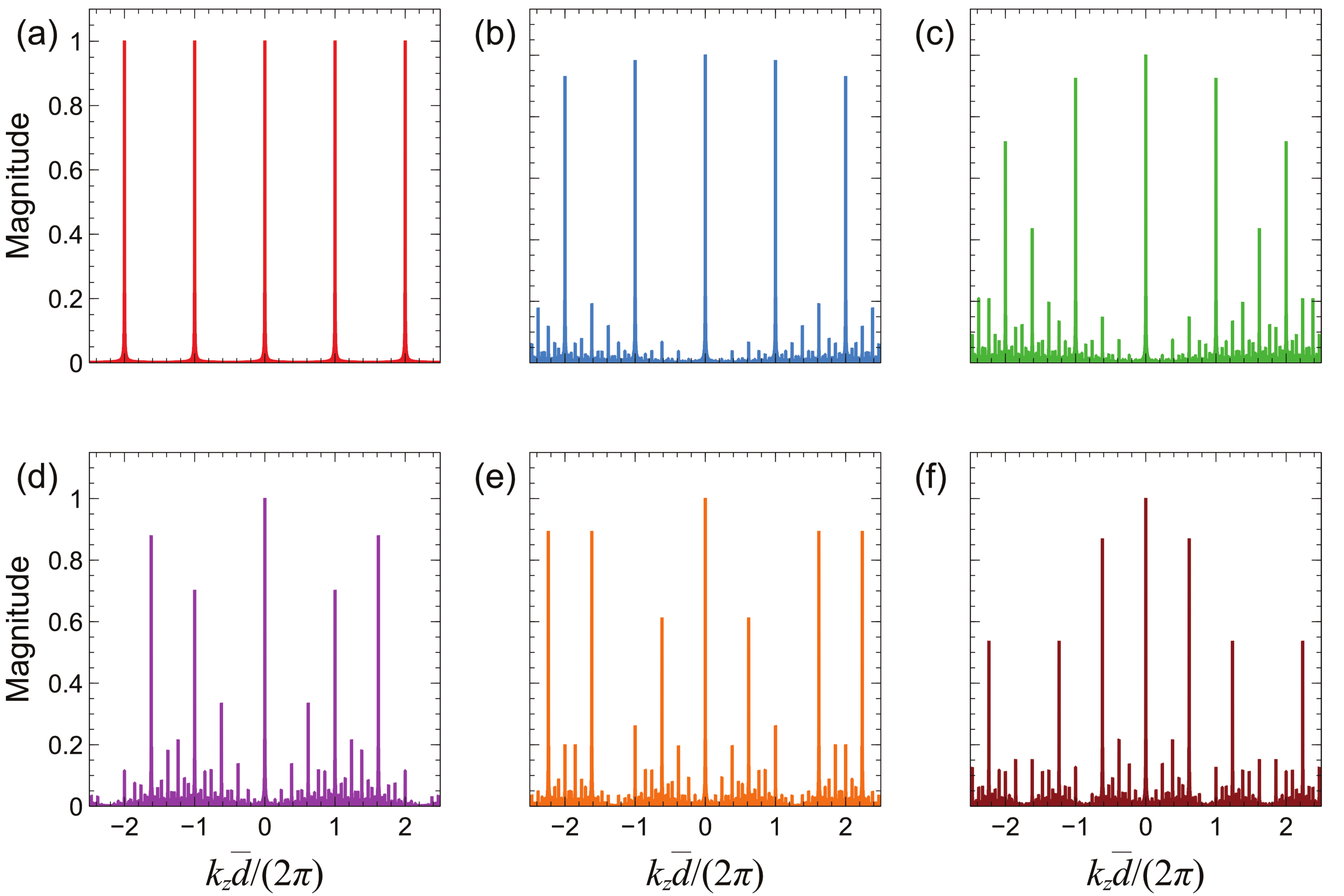}
	\caption{Representative spatial spectra (magnitude) pertaining to a geometry with $N=512$ elements, for representative values of the scale-ratio parameter. (a) $\nu=1$ , (b) $\nu=0.9$, (c) $\nu=0.8$, (d) $\nu=1/\varphi$, (e) $\nu=0.4$, (f) $\nu=0.2$. The spectra are normalized with respect to the value at $k_z=0$.}
	\label{Figure2}
\end{figure}

For illustration, Fig. \ref{Figure2} shows some representative spatial spectra pertaining to a finite-size ($N=1024$) structure, for rational and irrational values of $\nu$. By focusing on the dominant spectral peaks, as $\nu$ decreases we observe a progressive weakening of the harmonics at integer values of $2\pi/\bar{d}$ (typical of periodicity) and the appearance of new dominant harmonics at intermediate positions.

The above modified-Fibonacci geometry has been studied in connection with antenna arrays \cite{Galdi:2005pq,Castaldi:2007rf} but, to the best of our knowledge, has never been applied to optical multilayers.

In all examples considered in our study below, the Fibonacci sequence is generated via (\ref{eq:Fibseq}), and the relative permittivity distribution starts with $\varepsilon_H$.

\subsection{Statement and Observables}
As shown in Fig. \ref{Figure1}, the structure under study is obliquely illuminated by a plane wave with transverse-electric (TE) polarization.
Specifically, we assume an implicit $\exp\left(-i\omega t\right)$ time-harmonic dependence, and a $y$-directed, unit-amplitude
electric field
\beq
E_y^{(i)}\left(x,z\right)=\exp\left[ik_e \left(x\sin\theta_i+z\cos\theta_i\right)\right],
\eeq
where $\theta_i$ is the angle of incidence, $k_e=k\sqrt{\varepsilon_e}$ is the ambient wavenumber in the exterior medium, and $k=\omega/c=2\pi/\lambda$ is the vacuum wavenumber (with $c$ and $\lambda$ denoting the corresponding speed of light and wavelength, respectively).

Starting from some pioneering experimental  \cite{Merlin:1985qp} and theoretical  \cite{Kohmoto:1987lo} studies in the 1980s, 
prior works on quasiperiodic Fibonacci-type multilayers have essentially focused on the diffractive regime of photonic quasicrystals ($d_{a,b}\lesssim\lambda$), 
and have elucidated the physical mechanisms underpinning the localization \cite{Gellermann:1994lo}, photonic dispersion \cite{Hattori:1994pd}, perfect transmission \cite{Huang:2001pt,Peng:2002si,Nava:2009pl}, bandgap  \cite{Kaliteevski:2001bs} and field-enhancement \cite{Hiltunen:2007mo} properties, as well as the
multifractal  \cite{Fujiwara:1989mw}, critical  \cite{Macia:1999pn} and band-edge states \cite{DalNegro:2003lt}. 
Besides the aforementioned differences in the geometrical model, a key aspect of our investigation is the focus on the {\em deeply subwavelength} regime $d_{a,b}\ll \lambda$. In this regime, for the assumed TE polarization, the optical response of the multilayer tends to be accurately modeled via conventional EMT in terms of an effective relative permittivity \cite{Sihvola:1999em}
\beq
{\bar \varepsilon}_{\parallel}=L^{-1}
\sum_{n=1}^{N}\varepsilon^{(n)}d^{(n)},
\label{eq:EMT}
\eeq
where $\varepsilon^{(n)}$ and $d^{(n)}$ represent the relative permittivity ($\varepsilon_{H,L}$) and thickness ($d_{a,b}$), respectively, of the $n$-th layer. For the modified-Fibonacci geometry under study, it can be shown (see Appendix \ref{Sec:AppB} for details) that the following approximation holds with good accuracy
\beq
{\bar \varepsilon}_{\parallel}\approx \frac{\varepsilon_H+\varepsilon_L}{2},
\label{eq:EMTapp}
\eeq
{\em irrespective} of the scale-ratio parameter. By virtue of this remarkable property, we can explore the transition from perfect periodicity to quasiperiodicity maintaining the same effective properties; in other words, by varying the scale ratio $\nu$, the multilayer maintains the same proportions of high- and low-permittivity constituents, so that the only difference is their spatial arrangement.

As we will show hereafter, contrary to conventional wisdom, the spatial order may play a key role also at deep subwavelength scales in co-action with mixed evanescent/propagating light transport. To elucidate this mechanism, we rely on a rigorous solution of the boundary-value problem based on the well-established transfer-matrix formalism \cite{Born:1999un} (see Appendix \ref{Sec:AppC} for details). Specifically, we calculate the transmission coefficient
\beq
\tau_N=\frac{\left.E_{y}^{(t)}\right|_{z=L}}{\left.E_{y}^{(i)}\right|_{z=0}}=\frac{2}{\chi_N+i\upsilon_N},
\label{eq:tauN}
\eeq
where $\chi_N$ and $\upsilon_N$ denote the {\em trace} and {\em anti-trace}, respectively, of the transfer matrix associated to a $N$-layer structure (see Appendix \ref{Sec:AppC} for details).
Other meaningful observables of interest are the reflection (and absorption, in the presence of losses) coefficient, as well as the field distribution in the multilayer.

\section{Representative Results}
\label{Sec:Results}

\subsection{Parametric Study}
To gain a comprehensive view of the phenomenology and identify the critical parameters, we carry out a parametric study of the transmission response of the multilayered metamaterial by varying the incidence direction, electrical thickness and number the layers, and scale ratio. In what follows we assume the same constitutive parameters for the layers ($\varepsilon_L=1$, $\varepsilon_H=5$) and exterior medium ($\varepsilon_e=4$) utilized in previous studies on periodic and aperiodic (either orderly or random)
geometries \cite{Sheinfux:2014sm,Sheinfux:2016cr,Sheinfux:2017oo,Castaldi:2018be,Coppolaro:2018ao,Coppolaro:2020eo}, so as to
facilitate direct comparison of the results. Recalling the approximation in (\ref{eq:EMTapp}), this corresponds to an effective medium with ${\bar \varepsilon}_{\parallel}\approx3$; we stress that this value is essentially independent of the scale ratio, and therefore holds for all examples considered in our study. In the same spirit, although we are not bound with specific sequence lengths, we assume power-of-two values for the number of layers $N$, similar to our previous studies on Thue-Morse \cite{Coppolaro:2018ao} and Golay-Rudin-Shapiro \cite{Coppolaro:2020eo} geometries.
Moreover, to ensure meaningful comparisons among different geometries, we parameterize the electrical thickness in terms of the average thickness $\bar{d}$ in (\ref{eq:dav}), so that structures with same number of layers have same electrical size. In order to maintain the average thickness for different values of the scale ratio, it readily follows from (\ref{eq:nu}) and (\ref{eq:dav}) that the layer thicknesses need to be adjusted as 
\beq
d_a=\frac{(1+\varphi)}{(\nu+\varphi)} \bar{d}, \quad d_b=\nu d_a.
\eeq

Our study below is focused on the deeply subwavelength regime $0.01\lambda<\bar{d}<0.05\lambda$, with incidence angles $30^o<\theta_i\lesssim 60^o$. This last condition implies that, for the assumed constitutive parameters, the field is evanescent in the low-permittivity layers and propagating in high-permittivity ones. Prior studies on periodic and aperiodic configurations \cite{Sheinfux:2014sm,Sheinfux:2016cr,Sheinfux:2017oo,Castaldi:2018be,Coppolaro:2018ao,Coppolaro:2020eo} have shown that the anomalous phase-accumulation mechanism underlying this mixed light-transport regime can induce a large amplification of the nonlocal effects, so that the optical response exhibits a strongly enhanced sensitivity to geometrical variations at deeply subwavelength scales. The maximum angle of incidence is chosen nearby the critical angle ${\bar \theta}_c=\arcsin\left(\sqrt{{\bar \varepsilon}_{\parallel}/\varepsilon_e}\right)\approx60^o$, which defines the effective-medium total-internal-reflection condition.

\begin{figure}
	\centering
	\includegraphics[width=16cm]{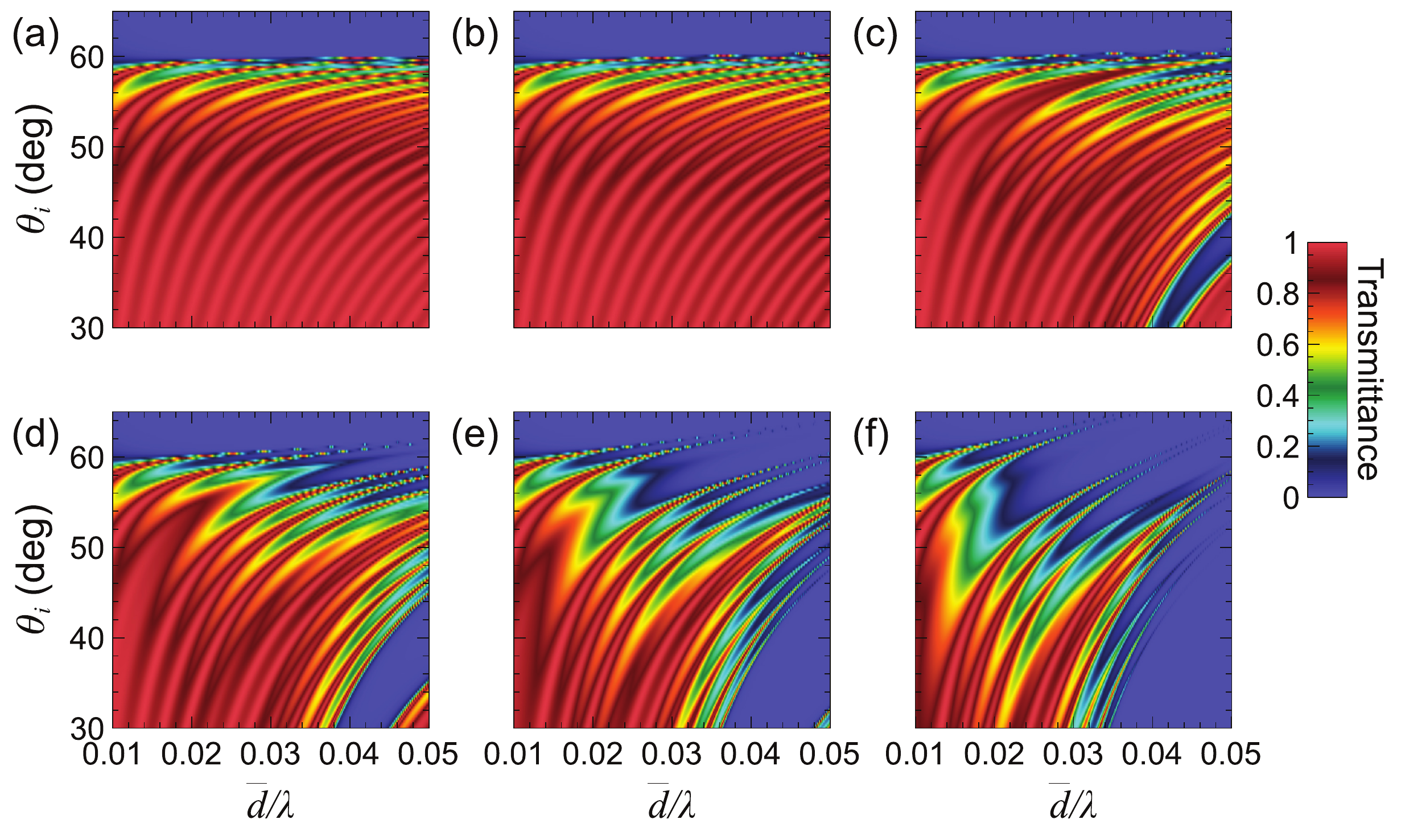}
	\caption{Comparison among the transmittance responses ($\left|\tau_N\right|^2$, in false-color scale) of multilayered dielectric metamaterials with modified-Fibonacci geometry, for $N=128$ layers, $\varepsilon_L=1$, $\varepsilon_H=5$, $\varepsilon_e=4$, as a function of the layer electrical thickness $d/\lambda$ and angle of incidence $\theta_i$, and for varying degrees of quasiperiodicity. (a) EMT prediction. (b), (c), (d), (e), (f) Responses for $\nu=1$ (perfect periodicity), $\nu=0.8$, $\nu=1/\varphi$, $\nu=0.4$, and $\nu=0.1$, respectively.}
	\label{Figure3}
\end{figure}

\begin{figure}
	\centering
	\includegraphics[width=16cm]{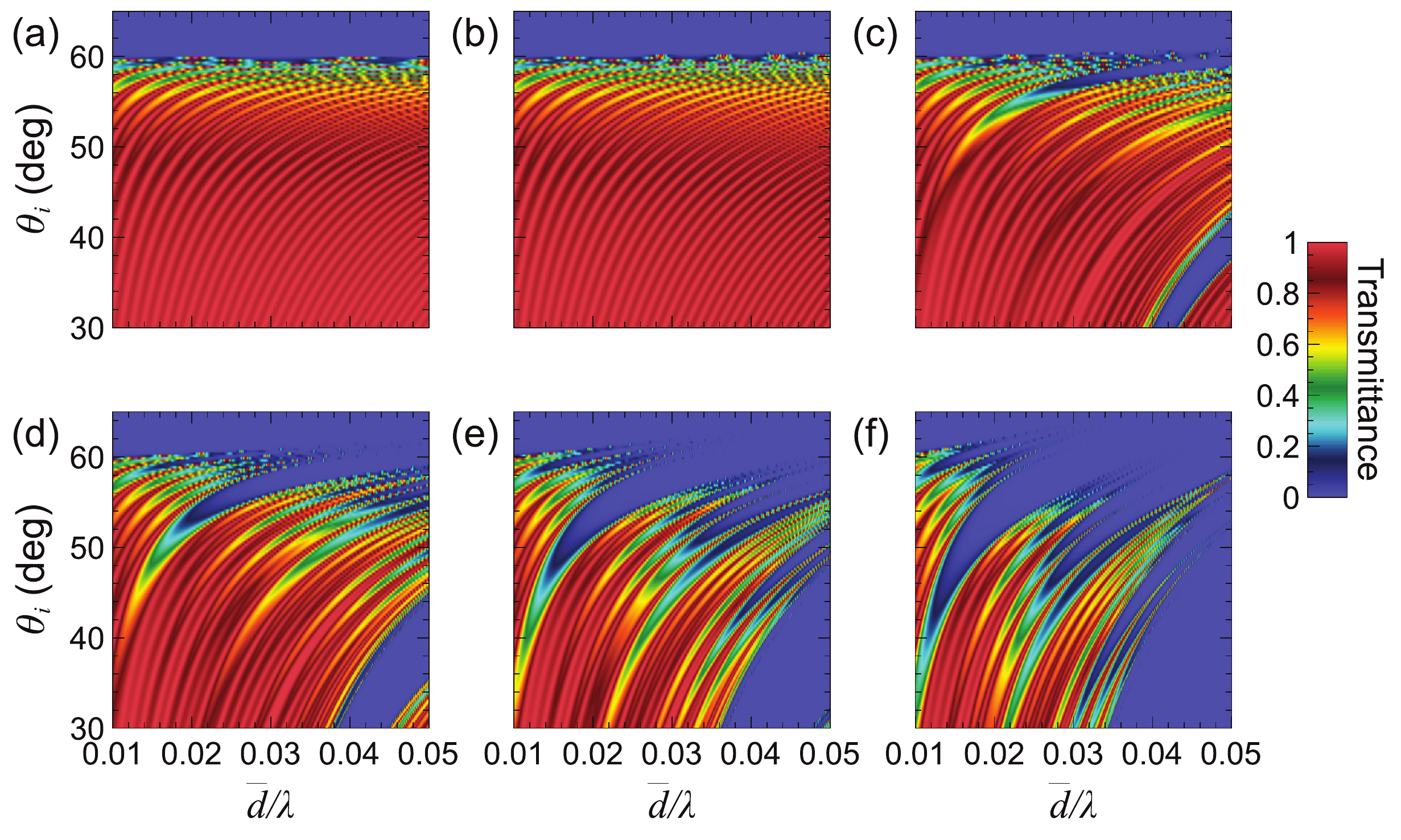}
	\caption{Comparison among the transmittance responses ($\left|\tau_N\right|^2$, in false-color scale) of multilayered dielectric metamaterials with modified-Fibonacci geometry, for $N=256$ layers, $\varepsilon_L=1$, $\varepsilon_H=5$, $\varepsilon_e=4$, as a function of the layer electrical thickness $d/\lambda$ and angle of incidence $\theta_i$, and for varying degrees of quasiperiodicity. (a) EMT prediction. (b), (c), (d), (e), (f) Responses for $\nu=1$ (perfect periodicity), $\nu=0.8$, $\nu=1/\varphi$, $\nu=0.4$, and $\nu=0.1$, respectively.}
	\label{Figure4}
\end{figure}

\begin{figure}
	\centering
	\includegraphics[width=16cm]{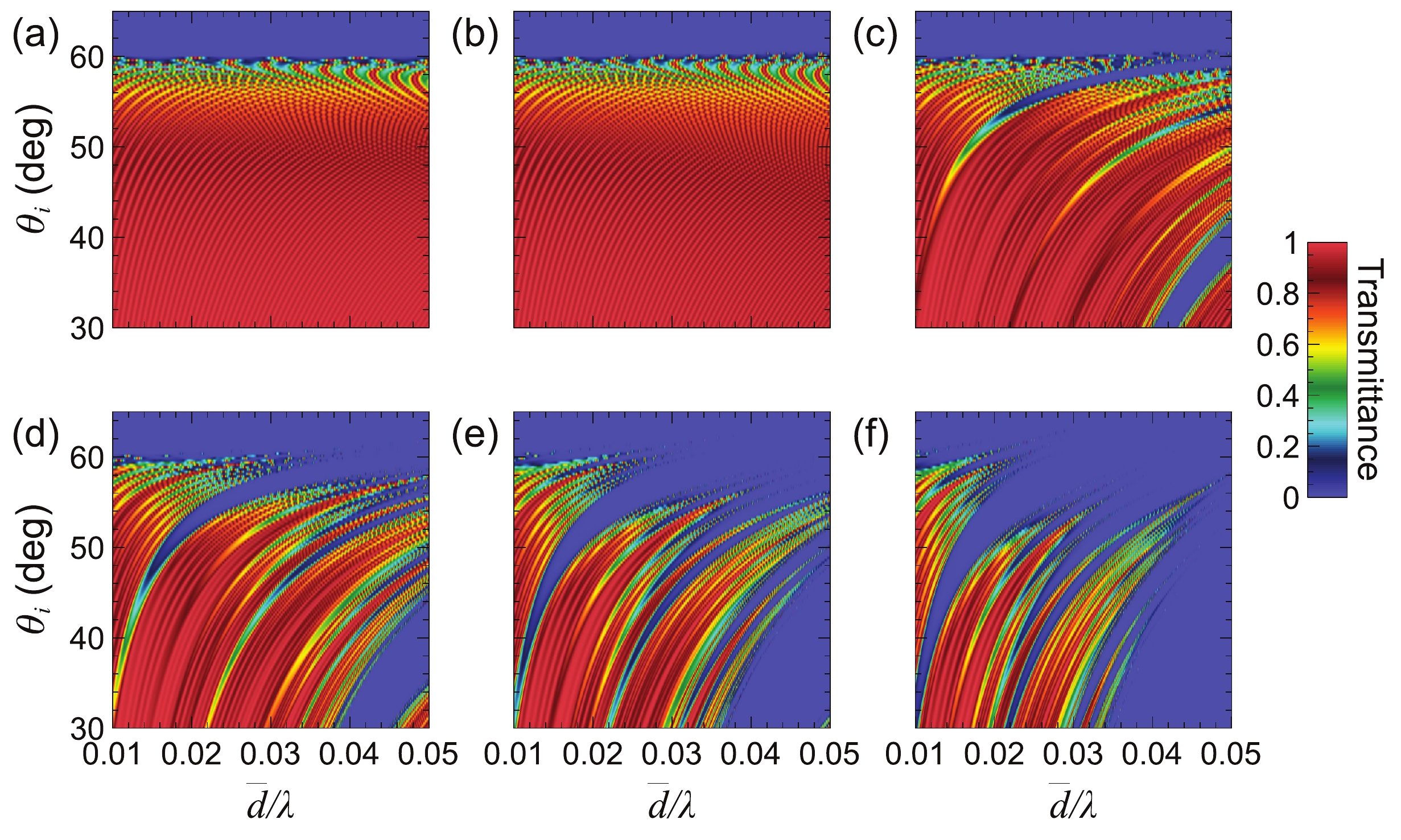}
	\caption{Comparison among the transmittance responses ($\left|\tau_N\right|^2$, in false-color scale) of multilayered dielectric metamaterials with modified-Fibonacci geometry, for $N=512$ layers, $\varepsilon_L=1$, $\varepsilon_H=5$, $\varepsilon_e=4$, as a function of the layer electrical thickness $d/\lambda$ and angle of incidence $\theta_i$, and for varying degrees of quasiperiodicity. (a) EMT prediction. (b), (c), (d), (e), (f) Responses for $\nu=1$ (perfect periodicity), $\nu=0.8$, $\nu=1/\varphi$, $\nu=0.4$, and $\nu=0.1$, respectively.}
	\label{Figure5}
\end{figure}

Figures \ref{Figure3}, \ref{Figure4} and \ref{Figure5} show the transmittance response ($\left|\tau_N\right|^2$) as a function of the average electrical thickness of the layers and angle of incidence, for $N=128, 256$ and $512$ layers, respectively. Each figure is organized in six panels, pertaining to representative values of the scale ratio transitioning from perfect periodicity ($\nu=1$) to different degrees of quasiperiodicity, with both rational ($\nu=0.8, 0.4, 0.2$) and irrational ($\nu=1/\varphi\approx 0.618$) values; also shown is the reference EMT response pertaining to the effective relative permittivity in (\ref{eq:EMTapp}).

At a qualitative glance, we observe a generally good agreement between the EMT and periodic configurations. As intuitively expected, both cases exhibit a regime of substantial transmission (with Fabry-P\'erot-type fringes) within most of the observation range, with an abrupt transition to opaqueness in the vicinity of the critical angle ${\bar \theta}_c\approx60^o$. Although it is somehow hidden by the graph scale, a closer look around the transition region would in fact reveal significant differences between the EMT and periodic responses, as extensively studied in \cite{Sheinfux:2014sm,Andryieuski:2015ae,Popov:2016oa,Lei:2017rt,Maurel:2018so,Castaldi:2018be,Gorlach:2020bc}. The quasiperiodic configurations display instead visible differences with the EMT and periodic counterparts, also away from the critical-incidence condition, manifested as the appearance of medium- and low-transmission regions whose extent and complex interleaving increases with increasing size and decreasing values of the scale-ratio parameter. In what follows, we carry out a systematic, quantitative analysis of these differences and investigate the underlying mechanisms.

\subsection{Near-Critical Incidence}
\begin{figure}
	\centering
	\includegraphics[width=12cm]{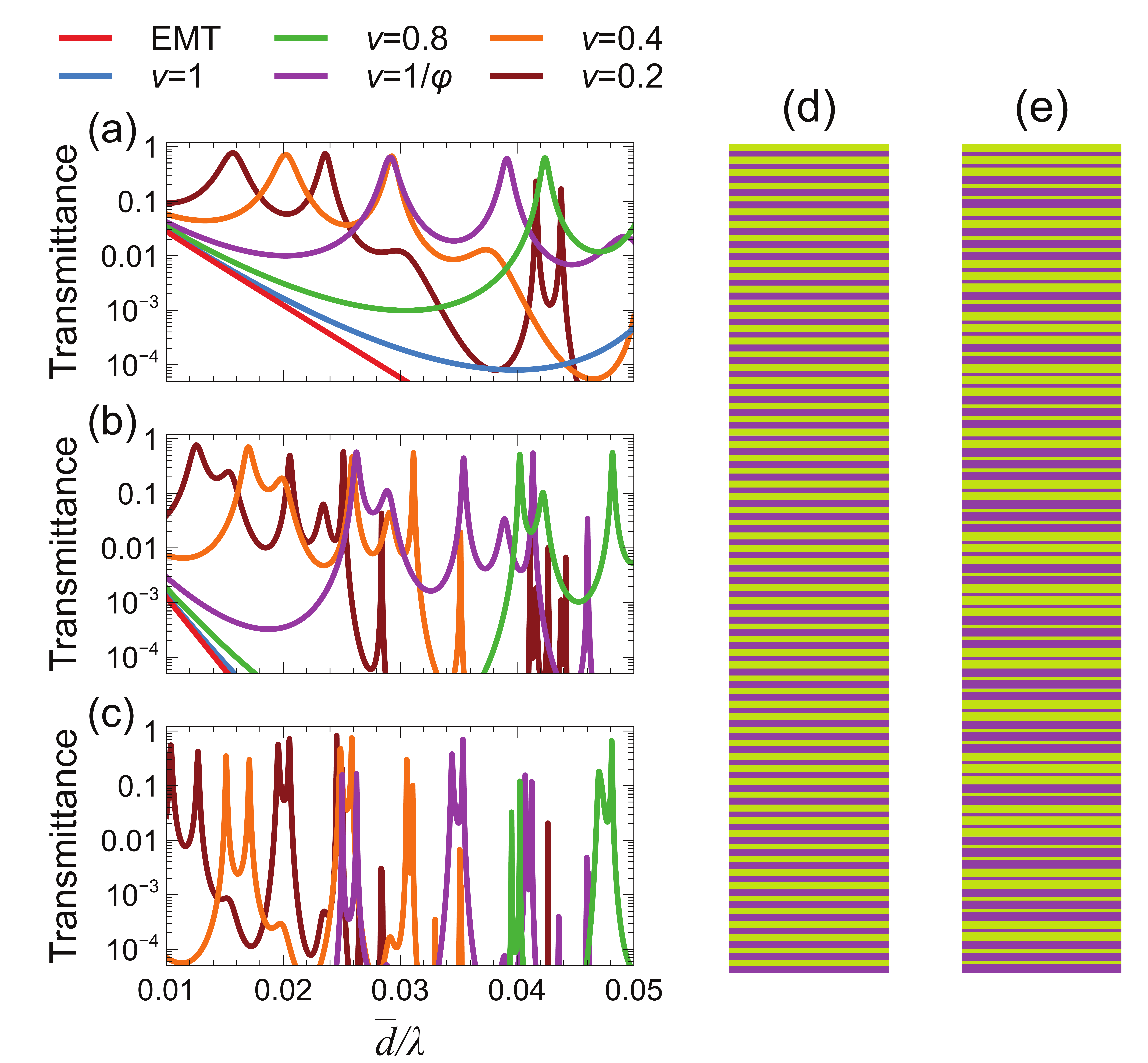}
	\caption{(a), (b), (c) Representative transmittance cuts from Figs. \ref{Figure3}--\ref{Figure5} at near-critical-incidence ($\theta_i=60.6^o$), for $N=128$, $256$, and $512$, respectively. Note the semi-log scale; in panel (c) the curves pertaining to the EMT and periodic cases are not visible since the transmittance level is below $10^{-4}$. (d), (e) Representative geometries for $N=128$ layers, with  $\nu=0.8$ and $\nu=0.4$, respectively; total lengths are not in scale.}
	\label{Figure6}
\end{figure}

As previously highlighted, nearby the critical angle $\theta_i\approx60^o$, substantial departures of the optical response from the EMT predictions can be observed also in the case of periodic geometries \cite{Sheinfux:2014sm,Andryieuski:2015ae,Popov:2016oa,Lei:2017rt,Maurel:2018so,Castaldi:2018be,Gorlach:2020bc}. However, the geometry under study exhibits different  types of anomalies that are distinctive of quasiperiodic order. As an illustrative example, Fig. \ref{Figure6} compares the transmittance cuts at $\theta_i=60.6^o$, for varying sizes and scale-ratios. For these parameters, the field in the EMT and periodic cases is evanescent and, although some differences are visible between the two responses, the transmission is consistently very low. Conversely, for increasing departures from periodicity, we start observing a general increase in the transmittance, with the occasional appearance of near-unit transmission peaks. As a general trend, for decreasing values of the scale-ratio parameter and increasing size, these peaks tend to narrow down, increase in number, and move toward smaller values of the electrical layer thickness. Perfect transmission peaks have been observed in previous studies on Fibonacci multilayers in the diffractive (quasicrystal) regime \cite{Huang:2001pt,Peng:2002si,Nava:2009pl}. From the theoretical viewpoint, they are a manifestation of extended optical states that can exist in quasiperiodic geometries as a consequence of enforced or hidden symmetries \cite{Nava:2009pl}. From the mathematical viewpoint, these peaks correspond to conditions where the trace of the transfer matrix is equal to two and the anti-trace vanishes [see (\ref{eq:tauN})]. 
Quite remarkably, in our case, these peaks may be observed even for electrical thicknesses as small as $\bar{d}\sim 0.01\lambda$, and relatively small ($N=128$) sizes. For basic illustration, Figs. \ref{Figure6}d and \ref{Figure6}e show two representative geometries associated with near-unit transmission peaks.

\begin{figure}
	\centering
	\includegraphics[width=16cm]{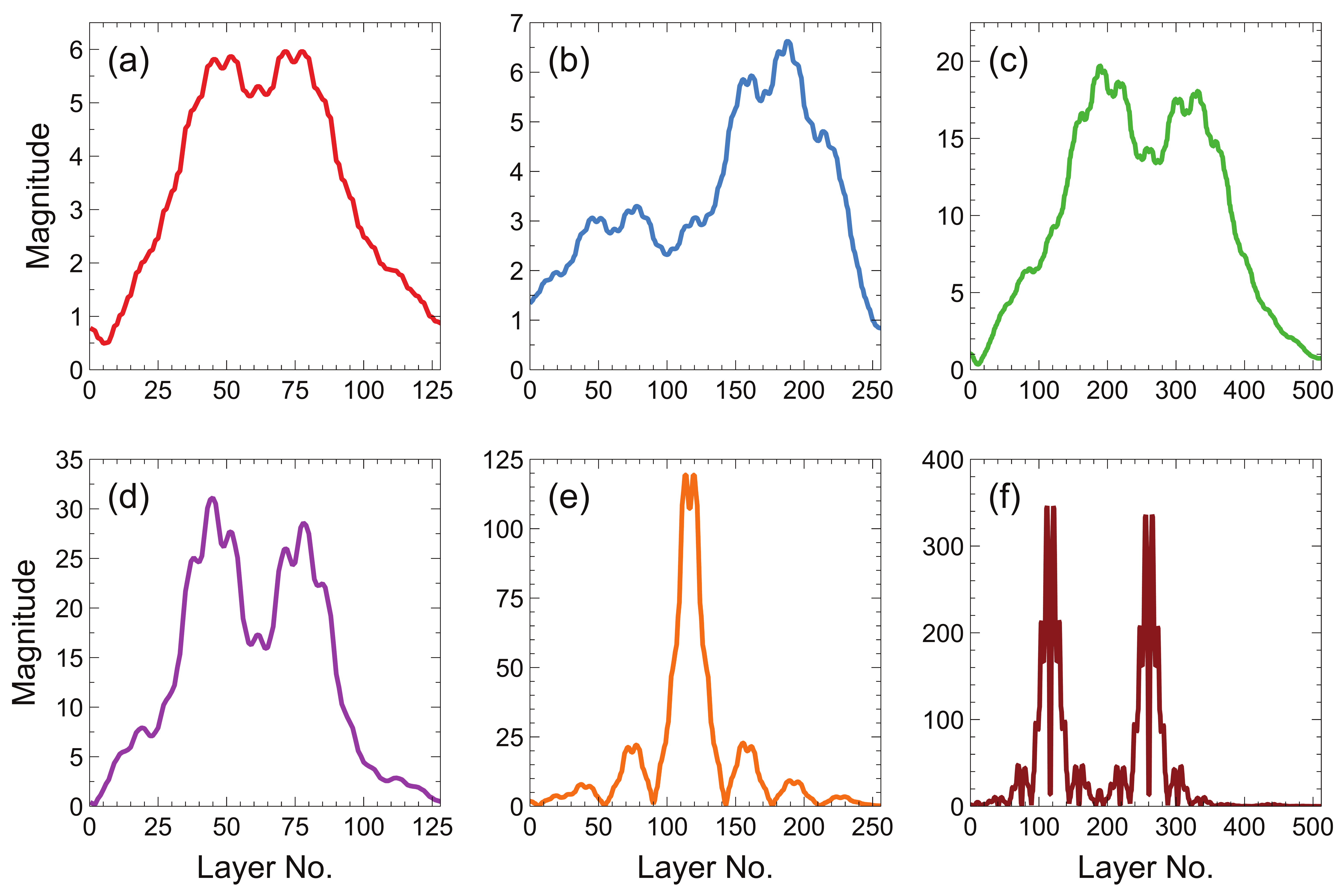}
	\caption{Representative electric-field (normalized-magnitude) distributions for near-critical-incidence states exhibiting high transmission and/or field enhancement. (a) $N=128$, $\nu=0.2$, $\bar{d}/\lambda=0.0157$, $\theta_i=60.6^o$. (b) $N=256$, $\nu=0.4$, $\bar{d}/\lambda=0.017$, $\theta_i=60.6^o$. 
	(c) $N=512$, $\nu=0.2$,  $\bar{d}/\lambda=0.0104$, $\theta_i=60.6^o$. (d) $N=128$, $\nu=1/\varphi$, $\bar{d}/\lambda=0.046$, $\theta_i=61.6^o$.
(e) $N=256$, $\nu=0.4$, $\bar{d}/\lambda=0.0366$, $\theta_i=60.85^o$. (f) $N=512$, $\nu=0.2$, $\bar{d}/\lambda=0.0478$, $\theta_i=59.35^o$.}
	\label{Figure7}
\end{figure}

Figures \ref{Figure7}a--\ref{Figure7}c show the field distributions (inside the multilayer) pertaining to three representative high-transmission peaks. Typical common features that can be observed include self-similarity and field enhancement; these characteristics have also been observed in the diffractive (photonic-quasicrystal) regime \cite{Fujiwara:1989mw,Hiltunen:2007mo}. In fact, for larger (but still deeply subwavelength) electrical thicknesses, field-enhancement factors up to $\sim 300$ can be observed for near-critical incidence, as exemplified in Figs. \ref{Figure7}d--\ref{Figure7}f.

\subsection{Non-Critical Incidence}
\begin{figure}
	\centering
	\includegraphics[width=8cm]{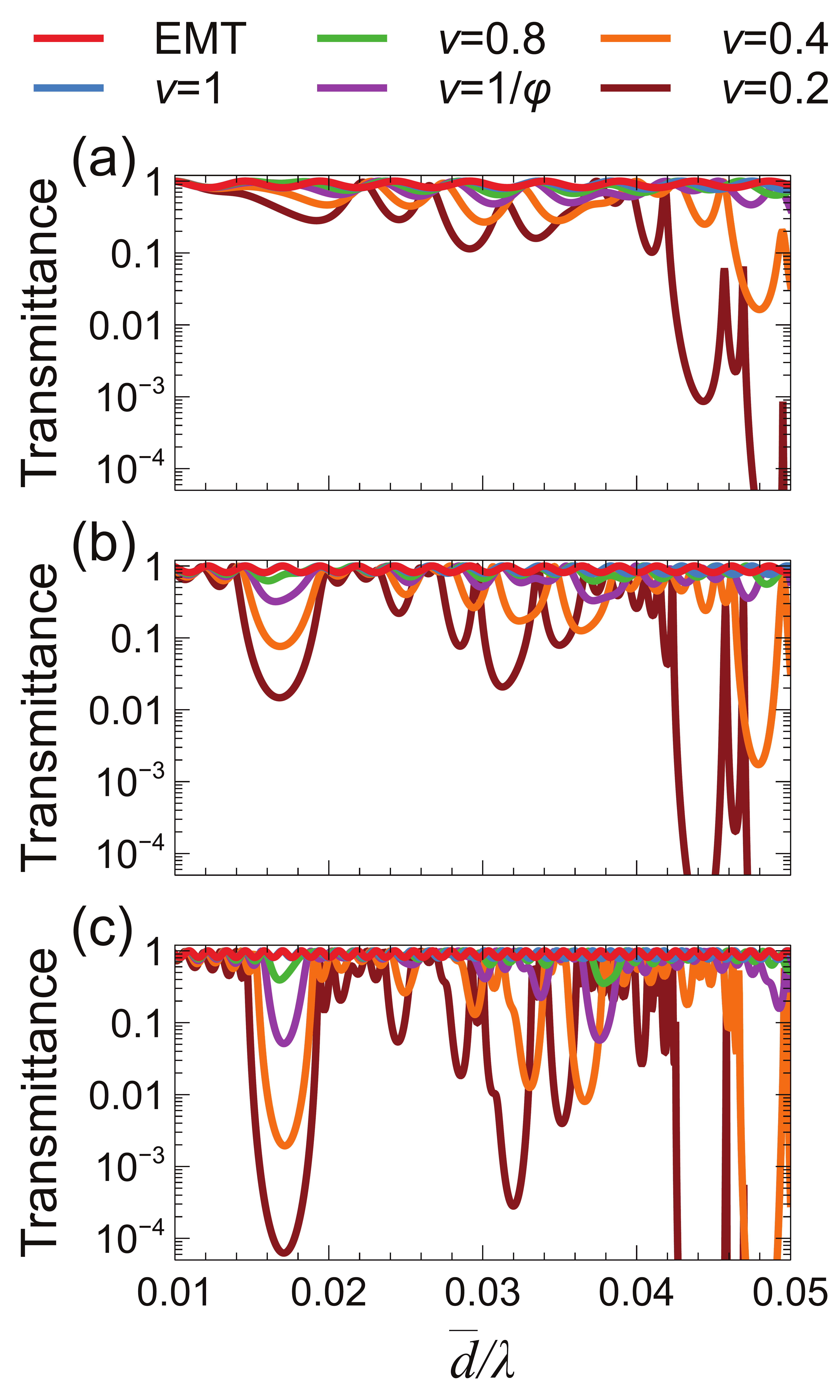}
	\caption{(a), (b), (c) Representative transmittance cuts from Figs. \ref{Figure3}--\ref{Figure5} away from critical incidence ($\theta_i=50.1^o$), for $N=128$, $256$, and $512$, respectively. Note the semi-log scale.}
	\label{Figure8}
\end{figure}

\begin{figure}
	\centering
	\includegraphics[width=8cm]{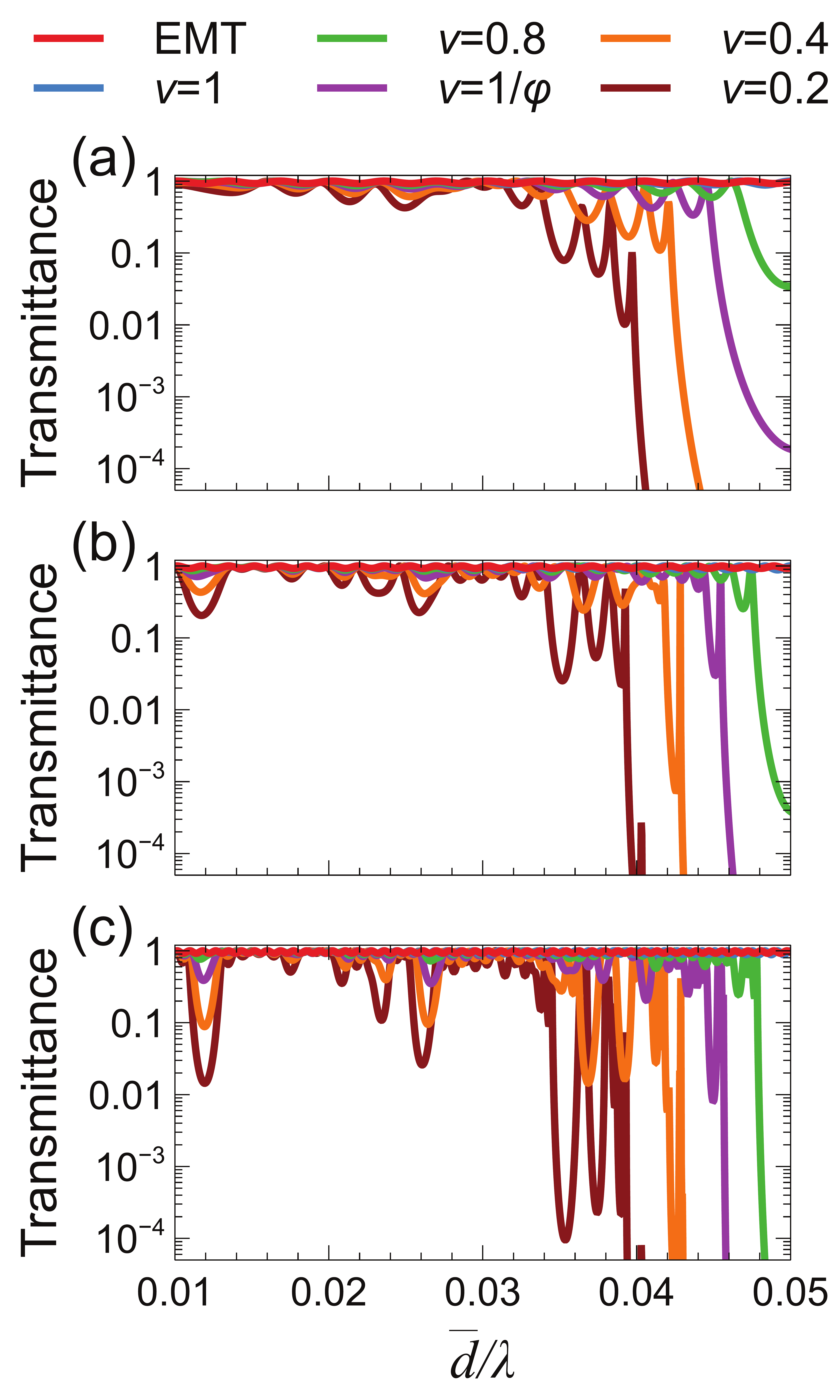}
	\caption{(a), (b), (c) Representative transmittance cuts from Figs. \ref{Figure3}--\ref{Figure5} away from critical incidence ($\theta_i=40.1^o$), for $N=128$, $256$, and $512$, respectively. Note the semi-log scale.}
	\label{Figure9}
\end{figure}

Away from critical-incidence conditions, the differences between the quasiperiodic and periodic/EMT configuration become even more pronounced. Figures \ref{Figure8} and \ref{Figure9} shows some representative transmittance cuts at $\theta_i=50.1^o$ and $40.1^o$, respectively. For these parameter configurations, the EMT and periodic responses are near-unit and hardly distinguishable. As the scale-ratio decreases, we observe the appearance of a rather wide bandgap at the upper edge of the electric-thickness range, and the progressive formation of secondary bandgaps at increasingly smaller values of the electrical thickness. For increasing sizes, these bandgaps tend to become denser and more pronounced. Quite interestingly, the position of certain bandgaps at particularly small values of the electrical thickness ($\bar{d}\sim 0.01\lambda$) seems to be rather robust with respect to the scale ratio.

\begin{figure}
	\centering
	\includegraphics[width=8cm]{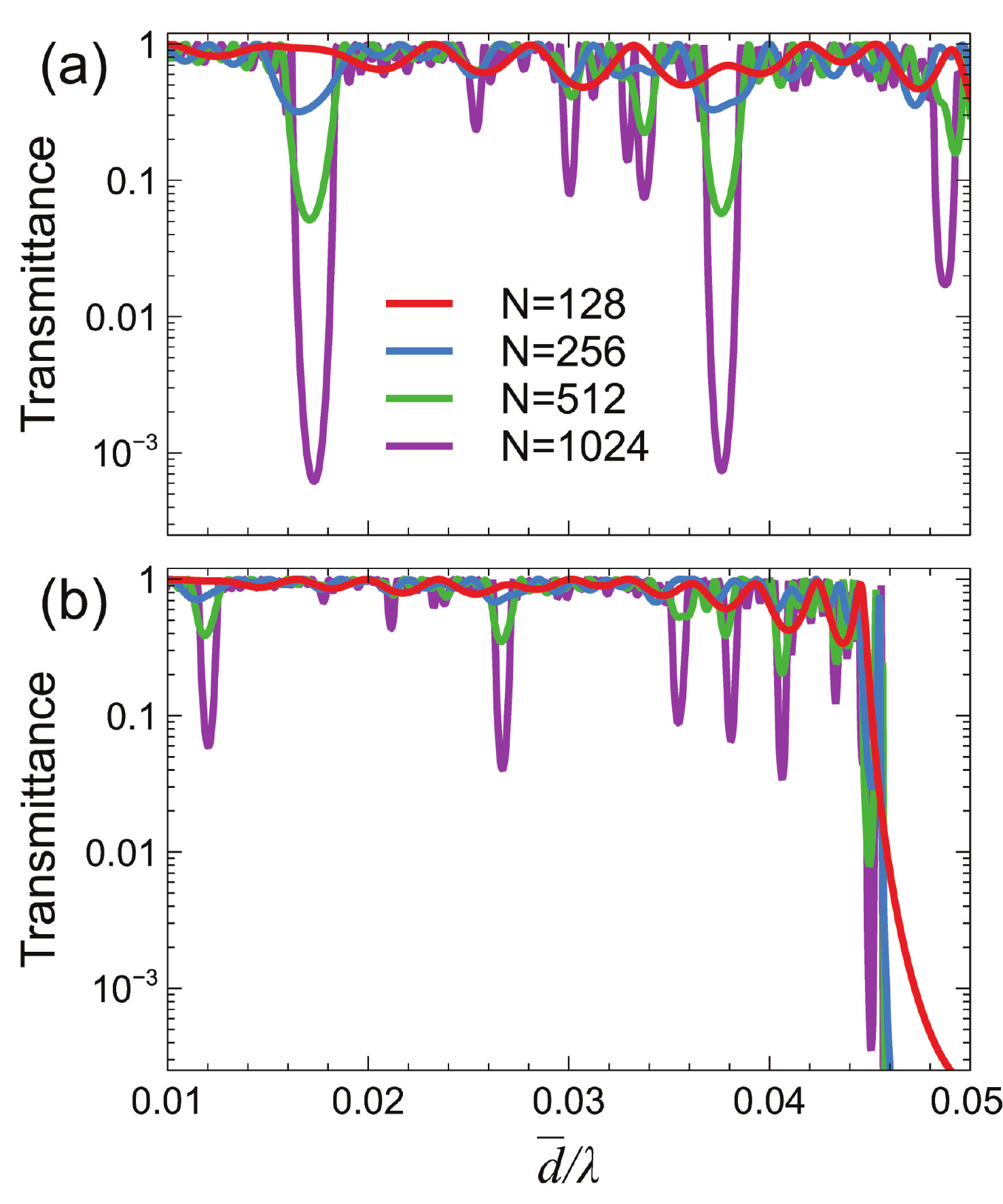}
	\caption{(a), (b) Representative transmittance cuts from Figs. \ref{Figure3}--\ref{Figure5} away from critical incidence ($\theta_i=50.1^o$ and $40.1^o$, respectively), for $\nu=1/\varphi$ and various values of the number of layers $N$. (a) Note the semi-log scale and the addition of the case $N=1024$.}
	\label{Figure10}
\end{figure}

To gain some insight in the effect of the structure size, Fig. \ref{Figure10} shows the transmittance cuts for a fixed value of the scale ratio ($\nu=1/\varphi$) for the number of layers $N$ ranging from 128 to 1024. As the size grows, we observe an increasing complexity with fractal-type structure. This is not surprising, as the fractal nature of the band structure is a well-known distinctive trait of Fibonacci-type photonic quasicrystals \cite{Kohmoto:1987lo,Kaliteevski:2001bs}, but it is still noteworthy that such complex behavior is visible at the deeply subwavelength scales of interest here.

\begin{figure}
	\centering
	\includegraphics[width=8cm]{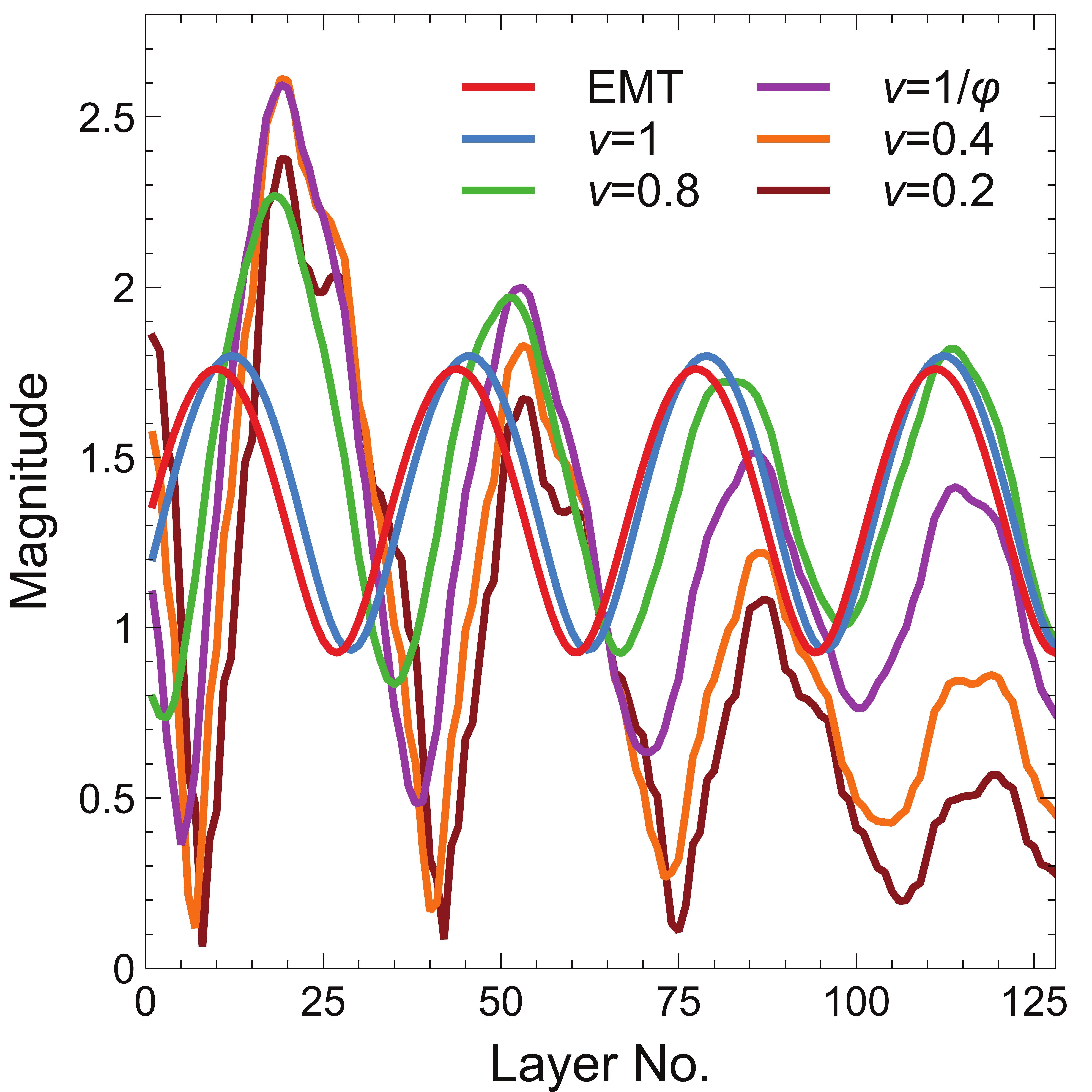}
	\caption{Comparison among electric-field (normalized-magnitude) distributions for $N=128$, $\bar{d}/\lambda=0.024$, and $\theta_i=54^o$, and various values of the scale-ratio parameter. Also shown, as a reference, is the EMT prediction.}
	\label{Figure11}
\end{figure}

To elucidate the role played by the scale ratio, Fig. \ref{Figure11} compares the field distributions for fixed size ($N=128$), non-critical incidence conditions ($\theta_i=54^o$) and electrical thickness $\bar{d}=0.024\lambda$, and various values of $\nu$. As can be observed, the field gradually transitions from a standing-wave, high-transmission character for the periodic case (in fair agreement with the EMT prediction), to a progressively decaying, low-transmission behavior as the scale ratio decreases. It is quite astounding that these marked differences emerge for layers as thin as $\bar{d}=0.024\lambda$ and a relatively small ($\sim 3\lambda$) structure. 

\begin{figure}
	\centering
	\includegraphics[width=16cm]{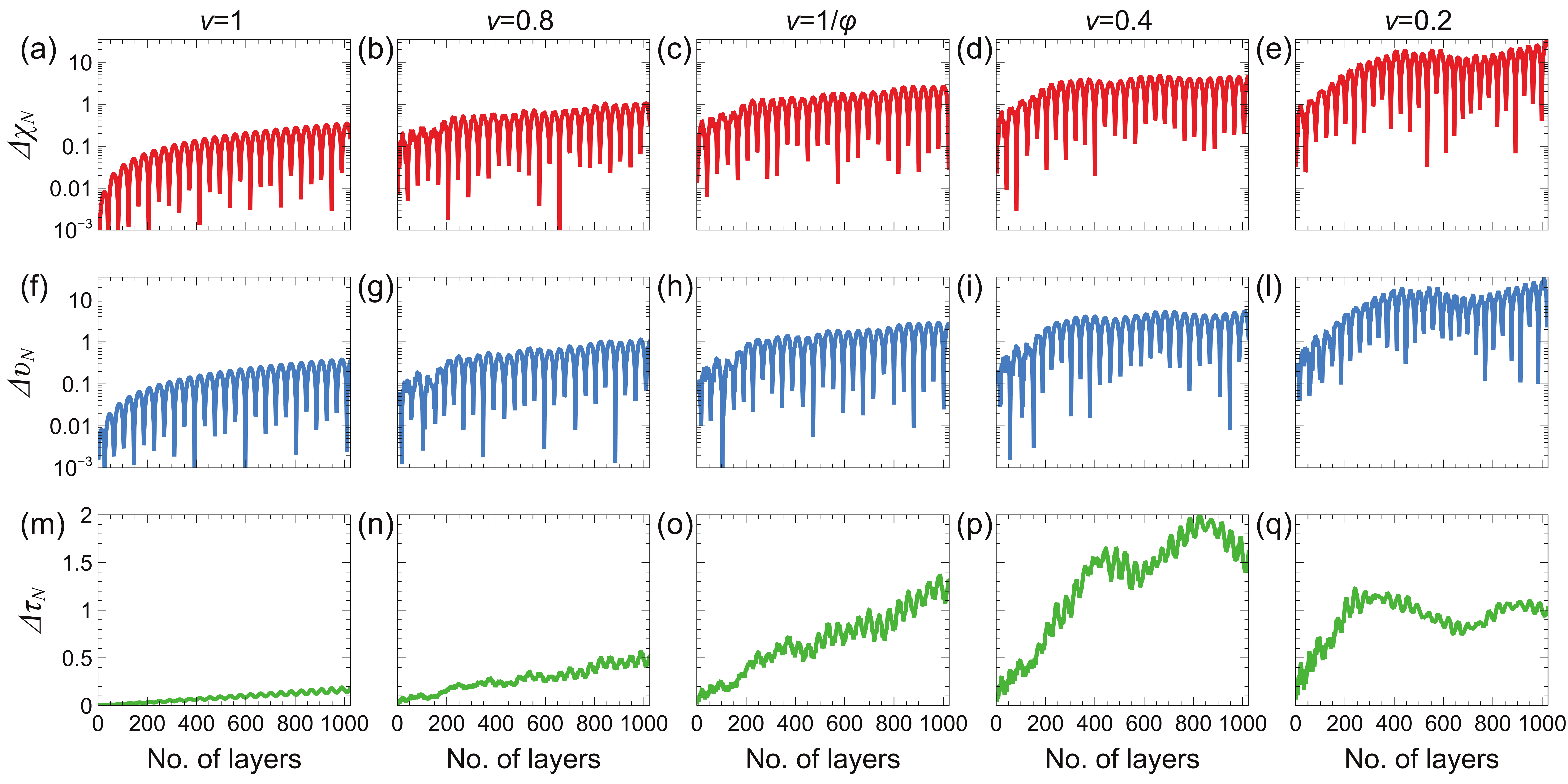}
	\caption{(a), (b), (c), (d), (e) Evolution of the trace error in (\ref{eq:errors}) as a function of the number of layers, for $\bar{d}/\lambda=0.015$ and $\theta_i=50^o$, and $\nu=1$, $0.8$, 1/$\varphi$, $0.4$, $0.2$, respectively. (f), (g), (h), (i), (l) Corresponding antitrace errors. (m), (n), (o), (p), (q) Corresponding transmission-coefficient errors. Note the semi-log scale in panels (a)--(l).}
	\label{Figure12}
\end{figure}

For the periodic \cite{Castaldi:2018be} and Thue-Morse \cite{Coppolaro:2018ao} geometries, it was shown that the EMT breakdown could be effectively interpreted and parameterized in terms of error propagation in the evolution of the trace and antitrace of the multilayer transfer-matrix, which is directly related to the transmission coefficient via (\ref{eq:tauN}). Interestingly, for the periodic case, it is possible to calculate analytically some closed-form bounds for the error propagation so as to identify the critical parameter regimes. Although for standard Fibonacci-type geometries (with both permittivity and thickness distributed according to the Fibonacci sequence) the trace and antitrace evolution can be studied via simple iterated maps  \cite{Kohmoto:1987lo,Wang:2000ta}, these unfortunately cannot be applied to our modified geometry. Nevertheless, they can be studied numerically from the transfer-matrix cascading (see Appendix \ref{Sec:AppC} for details). For $\theta_i=50^o$ and $\bar{d}=0.015\lambda$, Fig. \ref{Figure12} illustrates the evolution of the trace, antitrace and transmission-coefficient errors
\beq
\Delta\chi_N=\left|\chi_N-\bar{\chi}_N\right|,\quad
\Delta\upsilon_N=\left|\upsilon_N-\bar{\upsilon}_N\right|,\quad
\Delta\tau_N=\left|\tau_N-\bar{\tau}_N\right|,
\label{eq:errors}
\eeq
where the overbar indicates the EMT prediction; the evolution is shown as a function of the number of layers $N$, for representative values of the scale-ratio parameter. As a general trend, we observe fast, oscillatory behaviors with envelopes that grow with the multilayer size. For these parameters, the periodic case exhibits the slowest increase, with errors that remain below $\sim 0.1$; the reader is referred to Ref. \cite{Castaldi:2018be} for a detailed analytical study. As the geometry transitions to quasiperiodicity ($\nu<1$), we observe that the errors tend to grow increasingly faster with the number of layers, reaching values $\sim 10$ for the trace and antitrace, and approaching the maximum value of 2 for the transmission coefficient. These results quantitatively summarize at a glance the effects of quasiperiodicity in the EMT breakdown or, in other words, its visibility at deep subwavelength scales. Moreover, they also illustrate the important differences with respect to metallo-dielectric structures, which also feature a mixed (evanescent/propagating) light transport. In fact, for metallo-dielectric structures such as hyperbolic metamaterials, the errors in the trace and anti-trace can be significant even for a very small number of deeply subwavelength layers, thereby leading to visible ``bulk effects'', such as additional extraordinary waves \cite{Orlov:2011eo}. Conversely, in the fully dielectric case, the mechanism is essentially based on boundary effects, with errors that tend to be negligibly small for few layers, but, under certain critical conditions, may accumulate and grow (though non-monotonically) as the structure size increases.

\begin{figure}
	\centering
	\includegraphics[width=16cm]{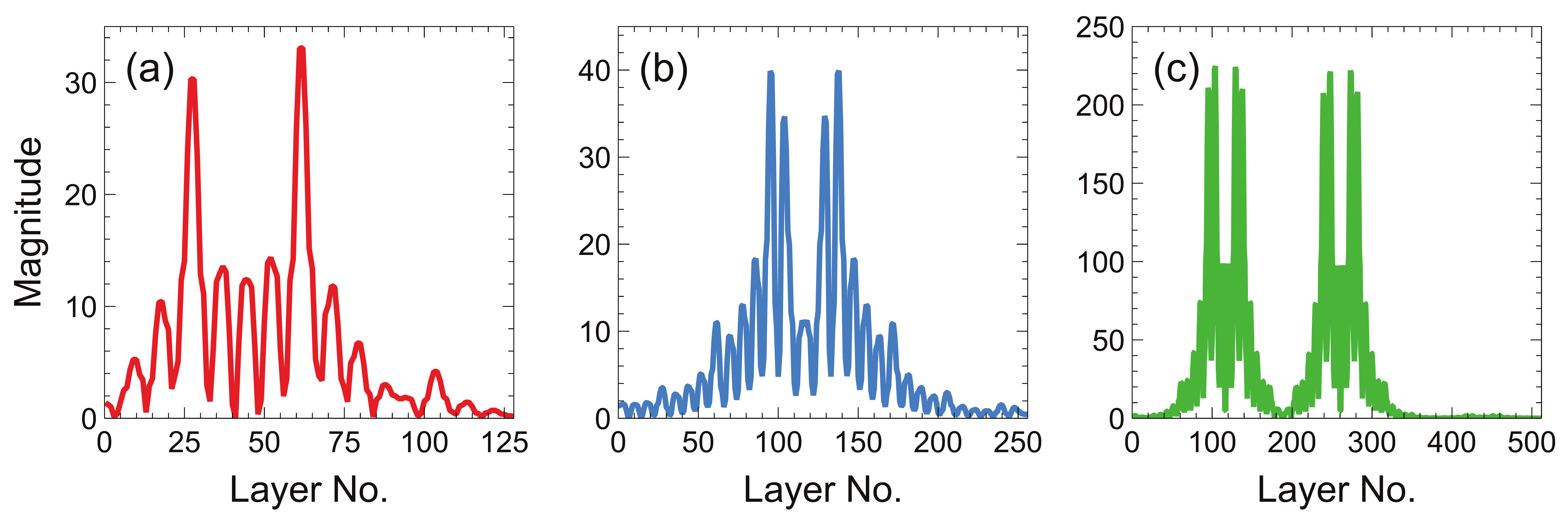}
	\caption{Representative electric-field (normalized-magnitude) distributions of states exhibiting high field enhancement.
		(a) $N=128$, $\nu=0.2$, $\bar{d}/\lambda=0.0498$, $\theta_i=52.35^o$.
		(b) $N=256$, $\nu=0.2$, $\bar{d}/\lambda=0.0443$, $\theta_i=46.1^o$.
	(c) $N=512$, $\nu=0.2$, $\bar{d}/\lambda=0.048$, $\theta_i=48.6^o$.}
	\label{Figure13}
\end{figure}

Strong field enhancement can also be observed for noncritical incidence. In this case, the most sensible enhancements are exhibited by edge modes around the bandgap appearing for $\bar{d}\sim 0.04\lambda$, still well within the deep subwavelength regime. Figure \ref{Figure13} illustrates three representative modes, for different sizes, scale ratio and incidence conditions.  For increasing size, we observe that the field distributions tend to exhibit self-similar, fractal-like structures, with enhancements of over two orders of magnitudes. Such levels of enhancement are in line what observed in prior studies on aperiodic geometries \cite{Coppolaro:2018ao,Coppolaro:2020eo} geometries, and in substantial contrast with the EMT prediction (see \cite{Coppolaro:2018ao} for details) 
\beq
\bar{\gamma} =\frac{\sqrt{\varepsilon_{e}} \cos \theta_{i}}{\sqrt{\bar{\varepsilon}_{\|}-\varepsilon_{e} \sin ^{2} \theta_{i}}},
\eeq
which, for the parameters in Fig. \ref{Figure13}, is $\lesssim 2$.

\subsection{Nonlocal Corrections}
\label{Sec:NL}
For the periodic case ($\nu=1$), it was shown \cite{Castaldi:2018be} that the error-propagation phenomenon illustrated in Fig. \ref{Figure12} could be significantly mitigated by resorting to suitable nonlocal corrections
(and possibly magneto-electric coupling \cite{Popov:2016oa}) in the effective-medium model, which could be  computed analytically in closed form. In principle, such strategy could be applied to the quasiperiodic scenario ($\nu<1$) of interest here, but there is no simple analytical expression for the nonlocal corrections. For a basic illustration, we resort to a fully numerical approach, by parameterizing the effective relative permittivity as
\beq
{\hat \varepsilon}_{\parallel}\left(k_x\right)=\frac{a_0 \left(1+a_2 k_x^2+a_4 k_x^4\right)}{1+b_2 k_x^2+b_4 k_x^4},
\label{eq:NLe}
\eeq
where the wavenumber dependence implies the nonlocal character (with only even powers of $k_x$ in view of the inherent symmetry), and the coefficients $a_0$, $a_2$, $a_4$, $b_2$, $b_4$ generally depend on the frequency and on the multilayer geometrical and constitutive parameters. These coefficients are computed numerically by minimizing the mismatch with the exact transmission response at selected wavenumber values (or, equivalently, incidence directions). Specifically, for a given multilayer and electrical thickness, we compute the coefficient $a_0$ by minimizing the mismatch for normal incidence ($k_x=0$), and the remaining four coefficients by minimizing the root-mean-square error for incidence angle $\theta_i$ varying from $1^o$ to $60^o$ (with step of $1^o$, and $k_x=k_e\sin\theta_i$). For the numerical optimization, we utilize a Python-based implementation of the Nelder-Mead method  available in the SciPy optimization library \cite{SciPy}.

\begin{figure}
	\centering
	\includegraphics[width=8cm]{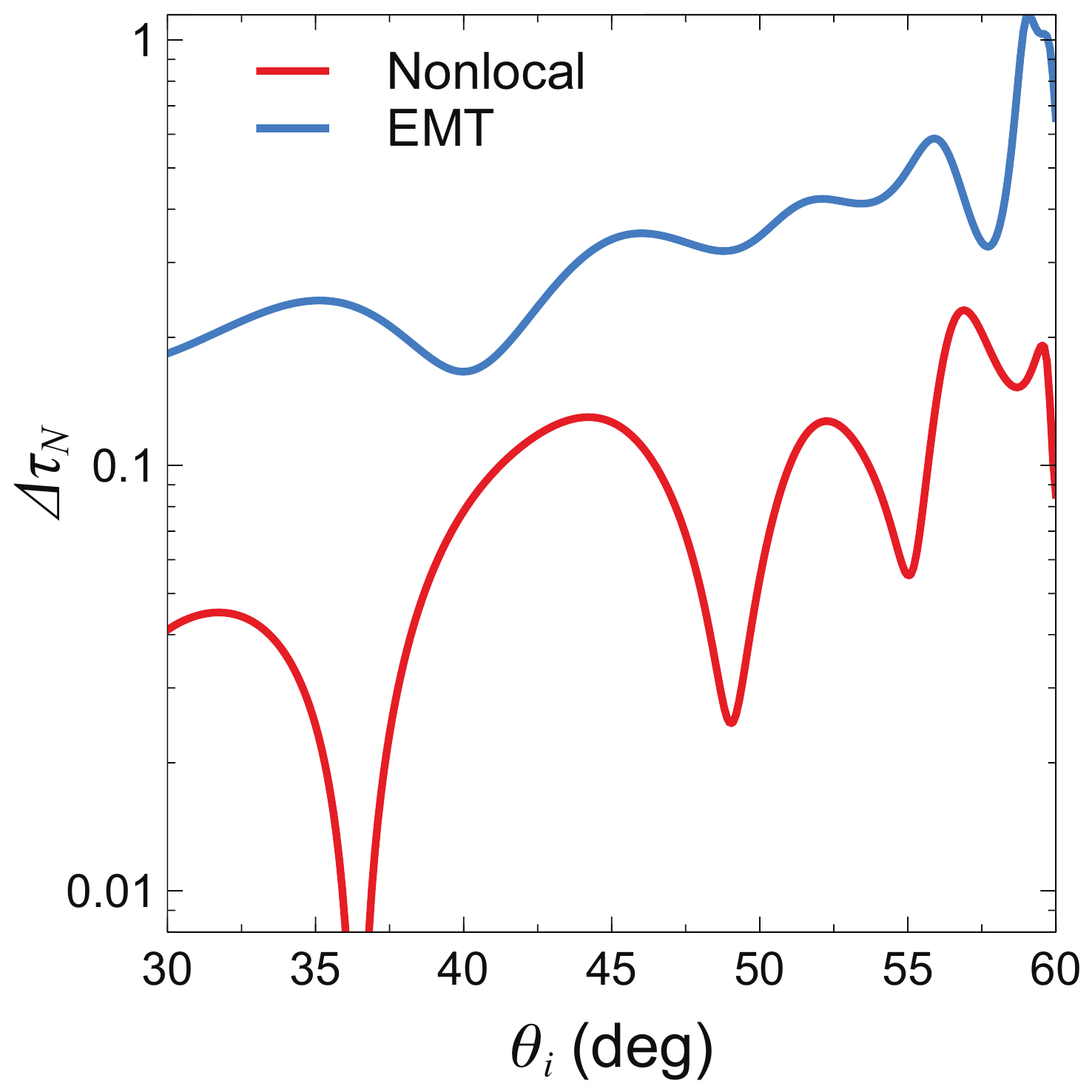}
	\caption{Transmission-coefficient  error $\Delta\tau_N$ in (\ref{eq:errors}) as a function of the incidence angle, for $N=128$, ${\bar d}=0.015\lambda$, considering the EMT prediction and the nonlocal effective model in (\ref{eq:NLe}) with parameters $a_0=3.043$, $a_2=0.242k_e^{-2}$, $a_4=-0.270k_e^{-4}$, $b_2=0.235k_e^{-2}$, $b_4=-0.267k_e^{-4}$. Note the semi-log scale.}
	\label{Figure14}
\end{figure}

Figure \ref{Figure14} illustrates some representative results, for $N=128$ layers, $\nu=0.4$ and ${\bar d}=0.015\lambda$. Specifically, we compare the transmission coefficient error $\Delta\tau_N$ in (\ref{eq:errors}) for the conventional EMT and the nonlocal effective model in (\ref{eq:NLe}) as a function of the incidence angle. As can be observed, a significant reduction is attained. Qualitatively similar results (not shown for brevity) are obtained for different lengths, frequencies and scale-ratio parameters. Stronger error reductions can be in principle obtained by resorting to higher-order and/or more sophisticated models that also account for magneto-electric coupling \cite{Popov:2016oa}.

\subsection{Anomalous Absorption and Lasing}
Our previous studies on the Thue-Morse \cite{Coppolaro:2018ao} and Golay-Rudin-Shapiro \cite{Coppolaro:2020eo} geometries have shown that, in the presence of small losses or gain, field-enhancement levels like those illustrated above can lead to anomalous absorption or lasing effects, respectively. To illustrate these phenomena, we assume a complex-valued
relative permittivity $\varepsilon_H=5+i\delta$, where the imaginary part $\delta$ parameterizes the presence of loss or gain (for $\delta>0$ and $\delta<0$, respectively, due to the assumed time-harmonic convention). 

\begin{figure}
	\centering
	\includegraphics[width=8cm]{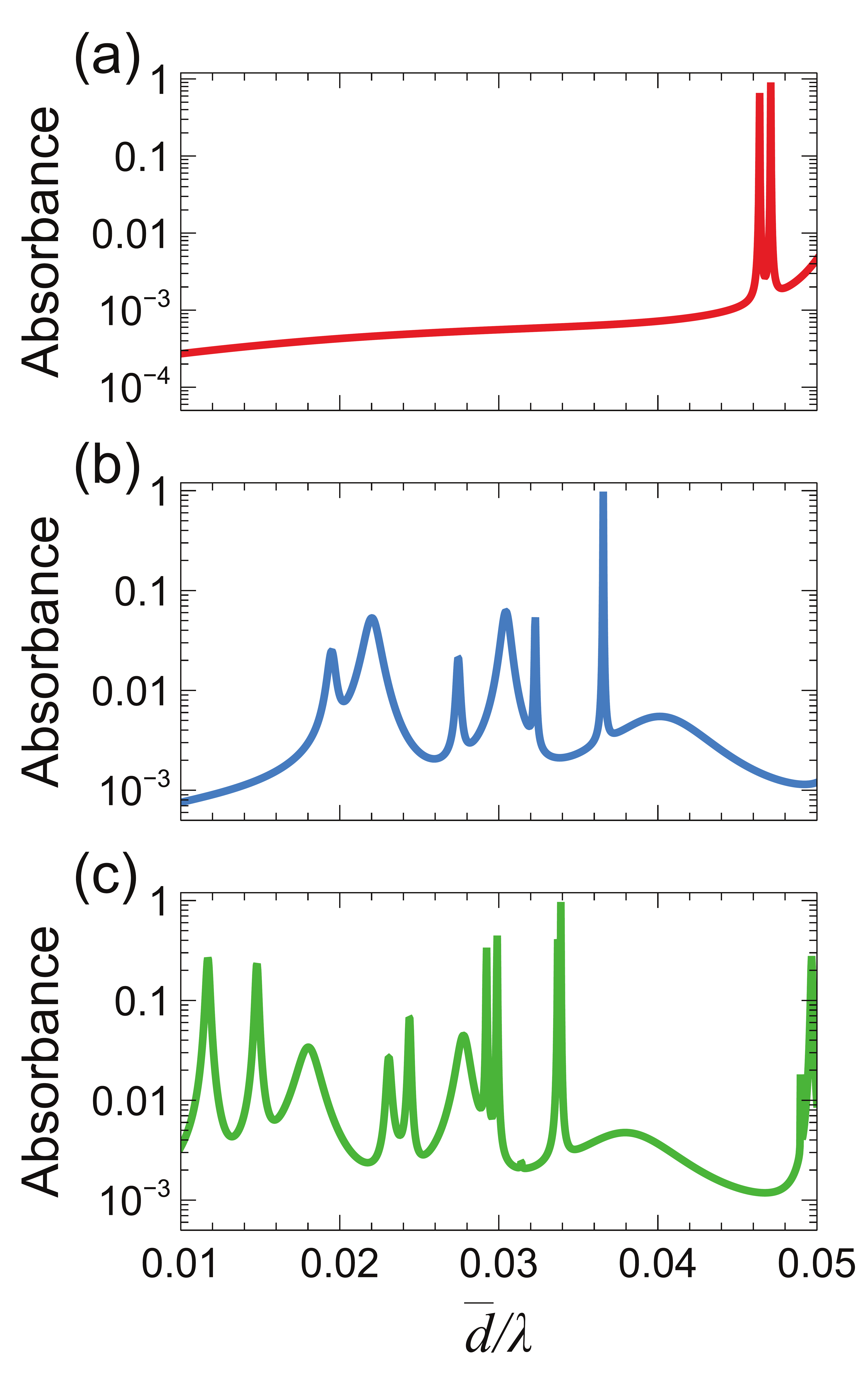}
	\caption{Representative absorbance responses, in the presence of small losses ($\varepsilon_H=5+i10^{-4}$), as a function of electrical thickness.
		(a) $N=128$, $\nu=0.2$, $\theta_i=67.1^o$.
		(b) $N=256$, $\nu=0.4$, $\theta_i=60.85^o$.
		(c) $N=512$, $\nu=0.4$, $\theta_i=60.35^o$. Note the semi-log scale.}
	\label{Figure15}
\end{figure}

For a very low level of losses ($\delta=10^{-4}$), Fig. \ref{Figure15} shows some representative absorbance responses, for different parameter configurations, from which we observe the presence of sharp peaks of significant (sometimes near-unit) amplitude. The corresponding field distributions (not shown for brevity) are qualitatively similar to those in Figs. \ref{Figure7} and \ref{Figure13}.
As a benchmark, for these parameters, the EMT prediction for the absorbance is $\lesssim 0.3$, whereas the result for the periodic reference configuration is $\lesssim 0.5$.

\begin{figure}
	\centering
	\includegraphics[width=8cm]{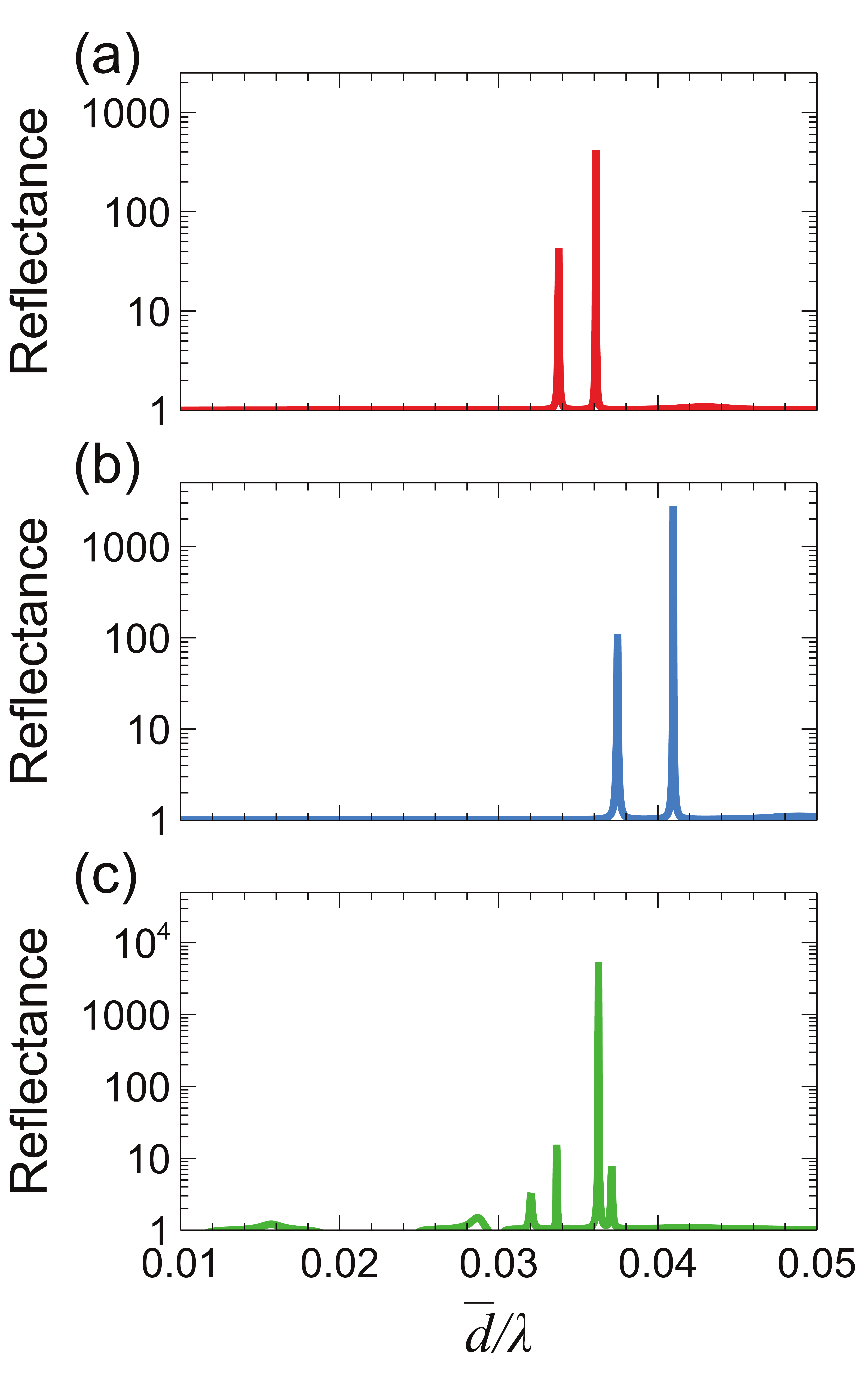}
	\caption{Representative reflectance responses, in the presence of small gain ($\varepsilon_H=5-i10^{-3}$), as a function of electrical thickness. (a) $N=128$, $\nu=0.2$, $\theta_i=64.35^o$.
		(b) $N=256$, $\nu=0.4$, $\theta_i=62.3^o$.
		(c) $N=512$, $\nu=1/\varphi$, $\theta_i=59.6^o$. Note the semi-log scale.}
	\label{Figure16}
\end{figure}

Finally, we consider the presence of small gain ($\delta=-10^{-3}$), and study the possible onset of lasing conditions.
Figure \ref{Figure16} shows some representative reflectance responses for different parameter configurations, which display sharp peaks with amplitude exceeding $\sim 1000$. This  indicates the presence of pole-type singularities that are distinctive of lasing, in spite of the quite low level of gain considered. To give an idea, by considering as a reference the lasing peak at $\bar{d}/\lambda=0.036$ in Fig. \ref{Figure16}a, in order to obtain comparable results in the EMT scenario we would need an increase of a factor $\sim 12$ in the gain coefficient or, equivalently, in the structure size (see \cite{Coppolaro:2018ao} for details).

These results provide further evidence of the potentially useful applications of
aperiodic order to the design of innovative absorbers and low-threshold lasers. 

\section{Conclusions and Outlook}
\label{Sec:Conclusions}
In summary, we have studied the effects of quasiperiodic order at deeply subwavelength scales in multilayered dielectric metamaterials. With specific reference to a modified-Fibonacci geometry, we have shown that the interplay with mixed evanescent/propagating light transport may induce anomalous optical responses (in terms of transmission, field-enhancement, absorption and lasing) that deviate substantially from the conventional EMT predictions. Moreover, by varying the scale-ratio parameter available in our model, we have explored and elucidated the transition from perfect periodicity to different shades of quasiperiodicity, identifying the critical parameter regimes and possible nonlocal corrections that can capture some of the effects. We highlight that, although our results here are restricted to TE polarization and a relatively high-contrast scenario, previous studies on the periodic case have shown that the EMT breakdown can also be observed for transverse-magnetic and/or lower-contrast configurations \cite{Zhukovsky:2015ed}, but their visibility may be reduced.

This investigation closes the loop with our previous studies on aperiodically (dis)ordered geometries, by adding to the already studied singular-continuous \cite{Coppolaro:2018ao} and absolutely continuous \cite{Coppolaro:2020eo} scenarios
a representative geometry with {\em discrete-spectrum} characteristics which had not been previously explored. These three characteristic spectra are fully representative of the generic aspects of aperiodic order.
Overall, these results indicate that deterministic spatial (dis)order may play a significant role even at deeply subwavelength scales. Besides providing a new geometrical degree of freedom in the design of optical devices (such as absorbers or lasers), this also opens up intriguing possibilities in the optical probing of the microstructure of a (meta)material and the sensing of its variations at scales much smaller than a wavelength.

Of particular interest for future studies it appears the exploration of similar effects 
 in non-Hermitian \cite{Dikopoltsev:2019co} and  time-varying \cite{Sharabi:2019lp} scenarios, 
 as well as the extension to 2-D geometries such as rod-type dielectric metamaterials.

\appendix

\section{Details on Spatial Spectrum}
Assuming two commensurate thicknesses $d_a$ and $d_b$, i.e., a rational scale ratio,
\label{Sec:AppA}
\beq
\nu=\frac{d_{b}}{d_{a}}=\frac{p_a}{p_b}, \quad p_a, p_b \in \mathbb{N},
\eeq
it readily follows from (\ref{eq:kzpq}) \cite{Buczek:2005pd} that
\beq
	k_{z\left(q_a+m p_a\right)\left(q_b+m p_b\right)}=k_{z q_a q_b}+m\left(\frac{2 \pi}{d_a} p_b\right),\quad m \in \mathbb{Z},
\eeq	
with $\mathbb{N}$ and $\mathbb{Z}$ denoting the sets of natural and integer numbers, respectively.
It then follows from (\ref{eq:Wpq}), that
\beq	
	W_{\left(q_a+m p_a\right)\left(q_b+m p_b\right)}=W_{q_a q_b},
\eeq
i.e., that the spatial spectrum is periodic with period $2 \pi p_b/{d_a}$. 

For the special case of a periodic structure ($d_a=d_b=d$, i.e.,  $\nu=1$), we obtain
\beq
\bar{d}=d, \quad W_{q_a q_b}=\left(p-q\right) \pi, \quad S_{pq}=\delta_{p q},
\eeq
with $\delta_{pq}$ denoting the Kronecker delta, thereby recovering the conventional spatial spectrum with peaks at $2\pi p/d$.

\section{Details on Eq. (\ref{eq:EMTapp})}
\label{Sec:AppB}
The result in (\ref{eq:EMTapp}) can be intuitively explained by recalling a well-know property of the Fibonacci sequences. It can be easily verified that, starting from the second iteration order of the inflation rules in (\ref{eq:itrule}), with the exception of the last two symbols, the Fibonacci sequence is {\em palindrome} \cite{Pirillo:1997fn}, i.e., it reads the same backward or forward. For instance, initializing the sequence with the symbol $a$, at the fifth iteration order we obtain $abaababaabaab$, which, omitting the last two symbols, yields $abaababaaba$, i.e., a palindrome.
It then readily follows that, for palindrome distributions of the thicknesses $d_a$ and $d_b$, and alternating distribution of the relative permittivities $\varepsilon_H$ and $\varepsilon_L$, the result in (\ref{eq:EMTapp}) holds {\em exactly}. In our case, we numerically verified that, for the assumed values of the sequence lengths and scale ratios, it provides a quite accurate approximation, with errors on the second decimal figure.

\section{Transfer-Matrix Formalism}
\label{Sec:AppC}
The tangential components of the electromagnetic field at the input and output interfaces of the generic $n$-th dielectric layer can be expressed as \cite{Born:1999un}
\beq
\left[\begin{array}{c}
	E_{y}^{(in)} \\
	i Z_{e} H_{x}^{(in)}
\end{array}\right]
={\underline{\underline {\cal M}}}_n
\cdot\left[\begin{array}{c}
	E_{y}^{(out)} \\
	i Z_{e} H_{x}^{(out)}
\end{array}\right],
\eeq
where 
\beq
Z_{e}=\frac{\omega \mu_{0}}{k_{z e}}
\eeq
is the wave impedance of the exterior medium for TE polarization, with $k_{ze}=k_e\cos\theta_{i}$ denoting the corresponding longitudinal wavenumber, and
 $\mu_0$ the vacuum permeability. Moreover,
\beq
\underline{\underline{{\cal M}}}_n=\left[\begin{array}{cc}
	\cos \left[k_z^{(n)} d^{(n)}\right] & \displaystyle{\frac{k_{z e}}{k_z^{(n)}}} \sin \left[k_z^{(n)} d^{(n)}\right] \\
	-\displaystyle{\frac{k_z^{(n)}}{k_{z e}}} \sin \left[k_z^{(n)} d^{(n)}\right] & \cos \left[k_z^{(n)} d^{(n)}\right]
\end{array}\right]
\label{eq:TM}
\eeq
is a unimodular transfer matrix \cite{Born:1999un}. In (\ref{eq:TM}),
\beq
k_z^{(n)}=k \sqrt{\varepsilon^{(n)}-\varepsilon_{e} \sin ^{2} \theta_{i}},
\eeq
is the local longitudinal wavenumber, and $\varepsilon^{(n)}$ and $d^{(n)}$ are the local relative permittivity ($\varepsilon_{H,L}$) and thickness ($d_{a,b}$), respectively.
Via chain multiplication of the transfer matrices of each layer, we can therefore obtain the transfer matrix of the entire multilayer \cite{Born:1999un}
\beq
\underline{\underline{{\cal M}}}=\prod_{n=1}^{N}
\underline{\underline{{\cal M}}}_n=
\left[\begin{array}{cc}
	{\cal M}_{11} & {\cal M}_{12} \\
	{\cal M}_{21} & {\cal M}_{22}
\end{array}\right].
\eeq
By expressing the input and output electric fields (for unit-amplitude incidence) in terms of the reflection and transmission coefficients ($\rho_N$ and $\tau_N$, respectively),
\begin{eqnarray}
E_{y}(x, z=0) &=&\left(1+\rho_{N}\right) \exp \left(i k x\sin\theta_i\right), \\
E_{y}(x, z=L) &=&\tau_{N} \exp \left(i k x\sin\theta_i\right),
\end{eqnarray}
and calculating the magnetic-field from the relevant Maxwell's curl equation,  we obtain the linear system
\beq
\left[\begin{array}{c}
	1+\rho_N \\
	-i\left(1-\rho_N\right)
\end{array}\right]=
\left[\begin{array}{cc}
	{\cal M}_{11} & {\cal M}_{12} \\
	{\cal M}_{21} & {\cal M}_{22}
\end{array}\right]
\cdot\left[\begin{array}{c}
	\tau_{N} \\
	-i \tau_{N}
\end{array}\right].
\label{eq:sys}
\eeq
From (\ref{eq:sys}), the expression in (\ref{eq:tauN}) follows straightforwardly by recalling the definitions of trace
\beq
\chi_N={\cal M}_{11}+{\cal M}_{22},
\eeq
and antitrace
\beq
\upsilon_N={\cal M}_{21}-{\cal M}_{12}
\eeq
of a matrix.


%

\newpage

%

%

%

%

%

%

%

%

%

%

%

%

%

%

%

%

\end{document}